\documentclass[a4paper,fleqn,usenatbib]{mnras}

\usepackage{newtxtext,newtxmath}
\usepackage[T1]{fontenc}
\usepackage{ae,aecompl}
\usepackage{bm}

\usepackage{graphicx}	% Including figure files

\usepackage{multirow}

\title[The source-count slope of FRBs]{The Slope of the Source-Count Distribution for Fast Radio Bursts}

\author[C. W. James et al.]{
C.\ W.\ James,$^{1,2}$\thanks{E-mail: clancy.james@curtin.edu.au (CWJ)}
R.\ D.\ Ekers,$^{1,3}$
J.-P.\ Macquart$^{1,2}$
K.\ W.\ Bannister,$^3$
and R.\ M.\ Shannon,$^{4}$
\\
$^1$International Centre for Radio Astronomy Research, Curtin University, Bentley, WA 6102, Australia\\
$^2$ARC Centre of Excellence for All-sky Astrophysics (CAASTRO), Australia\\
$^3$Australia Telescope National Facility, CSIRO Astronomy and Space Science, PO Box 76, Epping, NSW 1710, Australia\\
$^4$Centre for Astrophysics and Supercomputing, Swinburne University of Technology, PO Box 218, Hawthorn, VIC 3122, Australia
}%

\date{Accepted XXX. Received YYY; in original form ZZZ}

\pubyear{2018}

\begin{document}
\label{firstpage}
\pagerange{\pageref{firstpage}--\pageref{lastpage}}
\maketitle

\begin{abstract}
The slope of the source-count distribution of fast radio burst (FRB) fluences, $\alpha$, has been estimated using a variety of methods. Hampering all attempts have been the low number of detected FRBs, and the difficulty of defining a completeness threshold for FRB surveys. In this work, we extend maximum-likelihood methods for estimating $\alpha$, using detected and threshold signal-to-noise ratios applied to all FRBs in a sample without regard to a completeness threshold.
Using this method with FRBs detected by the Parkes radio telescope, we find $\alpha=-1.18 \pm 0.24$ (68\% confidence interval, C.I.), i.e.\ consistent with a non-evolving Euclidean distribution ($\alpha=-1.5$). Applying these methods to the Australian Square Kilometre Array Pathfinder (ASKAP) Commensal Real-time ASKAP Fast Transients (CRAFT) FRB survey finds $\alpha=-2.2 \pm 0.47$ (68\% C.I.). A full maximum-likelihood estimate finds an inconsistency with the Parkes rate with a p-value of 0.86\% ($2.6\, \sigma$). If not due to statistical fluctuations or biases in Parkes data, this is the first evidence for deviations from a pure power law in the integral source-count distribution of FRBs. It is consistent with a steepening of the integral source-count distribution in the fluence range 5--40\,Jy\,ms, for instance due to a cosmological population of FRB progenitors evolving more rapidly than the star-formation rate, and peaking in the redshift range 1--3.
\end{abstract}


\section{Introduction}
\label{sec:intro}

Fast radio bursts (FRBs) are one of the most-poorly understood astrophysical phenomena. FRBs are radio pulses with measured durations of milliseconds, arriving with high dispersion measures inconsistent with a Galactic origin. This makes their intrinsic luminosity extreme, with a plethora of models proposed to explain their progenitors.

The nature of FRB progenitors is so poorly constrained because, with the exception of the only known repeating source (FRB 121102; \citet{2016Natur.531..202S}), FRBs arrive from unpredictable directions at unpredictable times. Despite estimates for the all-sky daily rate being in the hundreds to thousands, the total number of detected FRBs is only approximately 50 (\citet{2016PASA...33...45P} (www.frbcat.org); \citet{craft_nature}).

Efforts to understand the distribution of FRB fluences tend to assume a power-law distribution for the all-sky rate $R$ of FRBs above a given fluence threshold $F$ of the form:
\begin{eqnarray}
R(F) & = & R_0 \left( \frac{F}{F_0}\right)^{\alpha}\mbox{  sky}^{-1}\,\mbox{day}^{-1} \label{eq:rate}
\end{eqnarray}
where $R_0$ is the rate at fluence threshold $F_0$. In the case that FRBs originate from the local universe --- as suggested by the proximity of FRB 121102 \citep{2017Natur.541...58C,2017ApJ...834L...7T} --- a value of $\alpha=-1.5$ is expected.

\citet{craft_nature} have recently shown a dispersion-brightness relation for the FRB population probed by the Parkes and Australian Square Kilometre Array Pathfinder (ASKAP) telescopes, suggesting this population originates at distances up to redshifts of $2$--$3$. For cosmological populations of radio sources, source evolution and $k$-corrections interact with the intrinsic source luminosity distribution to produce unique features in the measured source count distribution (see e.g.\ \citet{1996IAUS..175..547W}), and the same is therefore expected of the observed FRB fluence distribution \citep{2018MNRAS.tmp.1976M}. FRB data have been too sparse however to test more complicated models. The approach taken here is to search for the first hints of such structure by finding deviations from the simple power-law model for the observed fluence distribution.

Estimates of the parameters of the simple power-law model vary greatly. \citet{2016MNRAS.460L..30C} find $R_0=6^{+4}_{-3}\cdot 10^3$ FRBs sky$^{-1}$ day$^{-1}$ (95\% confidence interval, C.I.) based on the nine FRBs detected by the High Time Resolution Universe (HTRU) survey at Parkes above the fluence--duration range from 0.13\,Jy\,ms at 0.128\,ms to 5.9\,Jy\,ms over 262\,ms. However, this estimate did not include the non-detection of the HTRU mid-latitude survey \citep{2014ApJ...789L..26P}. \citet{2018MNRAS.475.1427B} use a larger sample of Parkes FRBs, finding a rate of $1.7^{+1.5}_{-0.9} \cdot 10^3$ FRBs sky$^{-1}$ day$^{-1}$ above a `fluence-complete' limit of $2$\,Jy\,ms. \citet{2016MNRAS.461..984O} estimate both the FRB rate and $\alpha$ using 15 FRBs from seven surveys, finding $-0.8 \ge \alpha \ge -1.7$ (95\% C.I.), with rates above $10^5$\,sky$^{-1}$\,day$^{-1}$. Importantly, \citet{2016MNRAS.461..984O} use the ratio of observed to threshold signal-to-noise values in their likelihood estimate, a parameter to which we shall return in section~\ref{sec:theory}.

These estimates effectively calculate $R$ by dividing the total number of bursts by the total observation time and sensitive solid angle, and hence fluctuate due to the low number of FRBs observed. In addition, as \citet{2018MNRAS.474.1900M} have noted, the effective detection threshold $F_{\rm eff}$ and solid angle $\Omega_{\rm eff}$ of an FRB search depends on $\alpha$, with small negative values of $\alpha$ (e.g.\ -1) favouring strong bursts detected far from beam centre, and large negative values (e.g.\ -3) favouring bursts detected near threshold close to beam centre. Without accounting for these effects, estimates of $R$ will be highly biased.

\citet{ME} account for the interaction between $R$ and $\alpha$ in using 19\,FRBs detected in the ASKAP Commensal Real-time ASKAP Fast Transients (CRAFT) survey \citep{craft_nature}, finding $R$ to be between 6.9 ($\alpha=-1.1$) and 23 FRBs\,sky$^{-1}$\,day$^{-1}$ ($\alpha=-3.0$), above effective thresholds of 92\,Jy\,ms and 36\,Jy\,ms respectively, at the survey time resolution of $1.2656$\,ms.

\citet{2017AJ....154..117L} perform a maximum-likelihood fit of equation~\ref{eq:rate} to FRBs detected by several telescopes, accounting for the effect of idealised beamshapes. They find $R=587^{+336}_{-315}$ above $1$\,Jy\,ms for $\alpha=-0.91 \pm 0.34$ (95\% C.I.). \citet{2016ApJ...830...75V} also use maximum-likelihoods, estimating $\alpha$ using two methods. The observed fraction of multiple- to single-beam detections of Parkes yields $-0.52>\alpha>-1.0$ (90\% C.I.), and combining both detections and non-detections of several other instruments finds $-0.32 > \alpha > -0.92$, with a combined constraint of $-0.5 > -\alpha > -0.9$.

The constraints set by the maximum-likelihood methods of both \citet{2016ApJ...830...75V} and \citet{2017AJ....154..117L}, which come from comparing multi-telescope data, are sensitive to unmodelled differences in system response to the (unknown) DM distribution, time duration, and frequency dependence of FRBs. Neither include the effects of radio-frequency interference on detection efficiency, and use analytic approximations to beam patterns, deviations from which have been shown by \citet{2018MNRAS.474.1900M} to greatly affect implied FRB properties. Furthermore, both analyses include the Lorimer burst, which should be excluded from statistical calculations on the grounds of discovery bias \citep{2018MNRAS.474.1900M}.

In summary, while the methods of both \citet{2016ApJ...830...75V} and \citet{2017AJ....154..117L} are analytically sound, their quantitative results should be subject to revision.

A more robust method to estimate $\alpha$ is to apply the maximum-likelihood method of \citet{1970ApJ...162..405C} to FRB samples from a single telescope. Both \citet{2018MNRAS.475.1427B} and \citet{2018MNRAS.474.1900M} do so to different samples of FRBs detected by Parkes, including data from the SUrvey for Pulsars and Extragalactic Radio Bursts (SUPERB), estimating $\alpha=-2.6^{+0.7}_{-1.3}$ and $-2.2^{+0.6}_{-1.2}$ respectively. It is therefore an outstanding question as to whether or not the differences in the values of $\alpha$ obtained with different methods are due to the pitfalls of making comparisons between different telescopes, a true change in the FRB spectral slope at different fluences, a statistical fluctuation, or some other effect.

This paper concerns estimates of $\alpha$ and the maximum-likelihood method of \citet{1970ApJ...162..405C} --- estimates of the absolute rate $R$ will be left to a future work. In section~\ref{sec:theory}, we extend the maximum-likelihood method of \citet{1970ApJ...162..405C}, formulating it in terms of measured and threshold signal-to-noise ratios as per \citet{2016MNRAS.461..984O}. Our formulation avoids all of the aforementioned uncertainties in beamshape and FRB properties, and renders the concept of a `completeness fluence' irrelevant. We outline the correct method for applying this updated maximum-likelihood test to FRB data in section~\ref{sec:application}. The new approach is both simpler, and allows the use of a greater proportion of detected FRBs, than previous applications. This gives it greater statistical power, and less bias. This method is then applied independently to both Parkes and ASKAP FRBs in section~\ref{sec:cpdata}, which have sufficient detections to provide independent meaningful estimates of $\alpha$. A full maximum-likelihood optimisation, accounting for errors, is performed in section~\ref{sec:full}, which allows the likelihood ratios to test whether or not the same power-law distribution can account for both samples. The discussion of section~\ref{sec:discussion} compares the resulting estimates of $\alpha$, both to each other, and to those of previous authors, and interprets the result in terms of the cosmological source evolution of FRB progenitors.

\section{Generalisation of the method of Crawford et al.}
\label{sec:theory}

The application of maximum likelihood methods to source-counting statistics is often referenced to \citet{1970ApJ...162..405C}, who discuss the problem in terms of a population of $N(S)$ sources with flux densities greater than some value $S$. It is assumed that $N(S)$ has the form of a power-law:
\begin{eqnarray}
N(S) & = & k S^{\alpha}, \label{eq:nks_orig}
\end{eqnarray}
where $k$ is a constant, and $\alpha$ the index. Note that \citet{1970ApJ...162..405C} use $-\alpha$ for the index, and $a$ for the estimator of $\alpha$. Using here $\alpha^*$ as the maximum-likelihood estimator of $\alpha$, $\alpha^*$ is then given by:
\begin{eqnarray}
\frac{1}{\alpha^*} & = & -\frac{1}{M} \sum_{i} \ln s_i^{\prime} \label{eq:crawford}
\end{eqnarray}
for an observation of $M$ sources with flux densities $S_i$, where $s_i=S_i/S_0$ for a detection threshold $S_0$. The standard deviation of the result for large $M$, for which $\alpha^* \sim \alpha$, is given by:
\begin{eqnarray}
\sigma_{\alpha}(\alpha) & \sim & \sigma_{\alpha}(\alpha^*) \nonumber \\
& = & \frac{M \alpha^*}{(M-1)(M-2)^{0.5}}. \label{eq:sigma_alpha}
\end{eqnarray}
The distribution of $\alpha^*$ given $\alpha$, $p(\alpha^*|\alpha)$, follows a gamma distribution:
\begin{eqnarray}
p(\alpha^*|\alpha) & = & \frac{(-\alpha)^M}{M!} \left( \frac{M}{-\alpha^*} \right)^{M+1} e^{-\alpha M/\alpha^*}. \label{eq:gamma_dist}
\end{eqnarray}
\citet{1970ApJ...162..405C} also note that while the estimate $(\alpha^*)^{-1}$ of $\alpha^{-1}$ obtained from equation (\ref{eq:crawford}) is unbiased, the estimate of $\alpha$, $\alpha^*$, is biased. An unbiased estimate, $\alpha^{\prime}$, can be found using:
\begin{eqnarray}
\alpha^{\prime} & = & \frac{M-1}{M} \alpha^*. \label{eq:unbiased}
\end{eqnarray}
Error estimates for $\alpha^{\prime}$ will also be modified by the same factor.

Equation \ref{eq:nks_orig} and \ref{eq:crawford} specify both the source distribution, and the detection threshold $S_0$, in terms of a single variable, $S$. However, the flux density threshold $S_0$, or equivalently a fluence threshold $F_0$, of a transient source will vary with event duration. In the case of FRBs, both dispersion measure, and source position in the beam, are often-discussed complicating factors. In the following section, we first demonstrate that equation \ref{eq:crawford} is applicable to an arbitrarily complicated distribution when $s$ is simply the observed signal-to-noise ratio S/N relative to a threshold, S/N$_{\rm th}$. We also formulate the problem from hereon in terms of fluence $F$, which is the standard used by the FRB community.

\subsection{Maximum likelihood methods in a multi-dimensional observation space}
\label{sec:derivation}

Consider a set of parameters $\bmath{\theta}$ covering both intrinsic source properties (e.g.\ dispersion measure) and observational effects (e.g.\ beamshape). Defining the relative source distribution of events within the space $\bmath{\theta}$ as $k(\bmath{\theta})$, and total $K$ such that:
\begin{eqnarray}
K & = & \int d\bmath{\theta} \, k(\bmath{\theta}), \label{eq:Kk}
\end{eqnarray}
we assume that $k$ is independent of fluence threshold $F$, so that the integral source-count distribution still has the form:
\begin{eqnarray}
N(F) & = & K \left(\frac{F}{F_0}\right)^{\alpha}. \label{eq:nks}
\end{eqnarray}
Note that we now normalise $F$ relative to some threshold $F_0$ (c.f.\ equation (\ref{eq:nks_orig})). While the use of number-counts $N$, rather than a rate $R$, is arbitrary, we do so as a reminder that $\bmath{\theta}$ can include time-dependent factors.

For this formulation, the probability of an event $p(\bmath{\theta},F)$ occurring in the range $d \bmath{\theta} d F$ given a single observation is:
\begin{eqnarray}
\frac{d p(F,\bmath{\theta})}{d\bmath{\theta} dF} & = & -\frac{k(\bmath{\theta})}{C} \frac{\alpha}{F_0} \left( \frac{F}{F_0} \right)^{\alpha-1}. \label{eq:pftheta}
\end{eqnarray}
The normalising constant $C$ is required to ensure that the probability density, when integrated over the entire parameter space, equates to unity. Considering events between a threshold $F_{\rm th}$ and maximum $F_m$, where (crucially for this method) both depend on $\bmath{\theta}$ in an arbitrary way to account for possible experimental sensitivities, $C$ becomes:
\begin{eqnarray}
C & = & \int d\bmath{\theta} \int_{\rm F_{\rm th}(\bmath{\theta})}^{F_{\rm m}(\bmath{\theta})} \frac{-d N(F,\bmath{\theta})}{d F} dF \nonumber \\
& = & \int d\bmath{\theta} k(\bmath{\theta}) \left[ \left(\frac{F_{\rm th}(\bmath{\theta})}{F_0}\right)^{\alpha} - \left(\frac{F_{\rm m}(\bmath{\theta})}{F_0}\right)^{\alpha} \right]. \label{eq:cnorm}
\end{eqnarray}
Defining relative fluence threshold $s$ and maximum $b$ similarly to \citet{1970ApJ...162..405C}:
\begin{eqnarray}
s & \equiv & \frac{F}{F_{\rm th}(\bmath{\theta})} \\
b & \equiv & \frac{F_{\rm m}}{F_{\rm th}(\bmath{\theta})}, \label{eq:relative}
\end{eqnarray}
equation \ref{eq:pftheta} and \ref{eq:cnorm} can be written in terms of $s$, using $dF = F_{\rm th} ds$:
\begin{eqnarray}
\frac{d p(s,\bmath{\theta})}{ d\bmath{\theta} ds} & = & -\frac{k(\bmath{\theta})}{C^{\prime}} \alpha s^{\alpha-1} F_{\rm th}^{\alpha}(\bmath{\theta}) \label{eq:pstheta} \\
C^{\prime} & = & \int d\bmath{\theta} k(\bmath{\theta}) F_{\rm th}^{\alpha} (\bmath{\theta}) \left[1-b^{\alpha}(\bmath{\theta})\right],
\end{eqnarray}
where a factor of $F_0^{\alpha}$ has disappeared into the new normalisation constant, $C^{\prime}$. Practically, this means that an experiment observing $N$ events need not be concerned about the absolute scale $F_0$ of the power-law distribution when calculating the relative probability of observing $s$ --- however, the absolute scale is critically important when estimating the absolute rate $R$, necessary to compare different experiments. A simple example would be two experiments with thresholds $F_0$ of $1$\,Jy\,ms and $10$\,Jy\,ms, where the latter would measure a rate $R$ of $10^{\alpha}$ less than the former.

In the case where the relative maximum fluence, $b$, is independent of $\theta$, the total probability of observing $s$ given a single observation becomes:
\begin{eqnarray}
p(s) & = & \int p(s,\bmath{\theta}) d\bmath{\theta} \nonumber \\
& = & -\alpha \frac{ s^{\alpha-1}}{1-b^{\alpha}} \frac{1}{C^{\prime\prime}} ds \int d\bmath{\theta} k_{\rm obs}(\bmath{\theta}) F_{\rm th}^{\alpha} (\bmath{\theta}) \nonumber \\
C^{\prime\prime} & = & \int d\bmath{\theta} k_{\rm obs}(\bmath{\theta}) F_{\rm th}^{\alpha} (\bmath{\theta}) \label{eq:cprimeprime}.
\end{eqnarray}
Noting that the normalisation constant $C^{\prime \prime}$ simply cancels out the integral, $p(s)$ reduces to:
\begin{eqnarray}
p(s) & = & -\alpha \frac{ s^{\alpha-1}}{1-b^{\alpha}} ds. \label{eq:ps}
\end{eqnarray}
This is exactly the formula derived by \citet{1970ApJ...162..405C} in their equation (6). All results presented in that work then apply, specifically that in the case where $b=\infty$, $\alpha$ can be calculated as per equation~(\ref{eq:crawford}). To repeat their analysis, maximising the log-likelihood $\mathcal{L}$ in terms of estimated probabilities $p_i$ for each event $i$ gives:
\begin{eqnarray}
\mathcal{L} & = & \log \prod_{i=1}^{N} p_i(s) \\
& = & \sum_{i=1}^{N} \left[ \log (-\alpha) +(\alpha-1) \log s_i - \log(1-b^{\alpha}) \right] \nonumber.
\end{eqnarray}
Differentiating with respect to $\alpha$ produces:
\begin{eqnarray}
\frac{\partial \mathcal{L}}{\partial \alpha} & = & \sum_{i=1}^{N} \left[ \alpha^{-1} + \log s_i +\frac{b^{\alpha} \log b}{1-b^{\alpha}} \right]. \label{eq:pre_alpha}
\end{eqnarray}
Setting this to zero to find the maximum with respect to $\alpha$ defines the estimator $\alpha^*$. Using $b=\infty$ produces:
\begin{eqnarray}
\frac{N}{\alpha^*} + \sum_{i=1}^{N} \log s_i  & = & 0 \nonumber \\
\implies \frac{1}{\alpha^*} & = & \frac{-1}{N} \sum_{i=1}^{N} \log s_i. \label{eq:alpha}
\end{eqnarray}
In summary, if one can calculate the ratio $s$ between signal strength and detection threshold $F_{\rm th}(\bmath{\theta})$ for each event, one can estimate the slope of the cumulative source-count distribution $\alpha$ \emph{without} knowing the complex dependencies of that threshold on event parameters $\bmath{\theta}$, or the distribution of true source events with that space $k(\bmath{\theta})$ --- or even what the source space $\bmath{\theta}$ is.

\section{Application to FRB searches}
\label{sec:application}

Fast radio burst searches currently suffer from several complicating effects. Their (generally) once-off nature means that most FRBs are poorly localised, and hence their true fluence $F$ is observed at reduced sensitivity $F_{\rm obs} = B F$, where $B$ is the value of the antenna beam pattern at which they are observed. Their variable duration --- itself a function of dispersion measure, scattering, and intrinsic pulse width --- means that longer bursts require a higher fluence to be detected. Radio-frequency interference (RFI) can result in a time-dependent threshold, while different surveys have vastly different sensitivities. The analysis algorithms used by each experiment have their own complicated response patterns. The particular shape of each FRB pulse in both time and frequency --- and how this is measured at the experimental time/frequency resolution --- also affects the detection probability \citep{2015MNRAS.447.2852K}. While the source distribution $k(\bmath{\theta})$ will obviously be independent of some of these parameters (e.g.\ local RFI, position in beam), all may affect the fluence sensitivity threshold $F_{\rm th}(\bmath{\theta})$.

It is important to emphasise that all these parameters fall within the definition of the parameter space $\bmath{\theta}$ introduced in an abstract sense in section~\ref{sec:theory}. Indeed, $\bmath{\theta}$ can extend to include different experiments in the one analysis. 

Correctly incorporating these experimental effects into an analysis will result in a powerful probe of the FRB fluence distribution. However, they make it very difficult to reconstruct the true value of $F$ for any given event. Furthermore, the intrinsic FRB distribution is related to the observed properties through equation~(\ref{eq:cnorm}), where in this example, the `normalising constant' $C$ is the total number of detected events.  Estimating the FRB rate therefore requires integrating over the parameter space $\bmath{\theta}$ of experimental dependencies. This is very difficult, since each telescope's $F_{\rm th}(\bmath{\theta})$ is often not well-understood, and the DM, duration, frequency, and signal-shape dependence of $k(\bmath{\theta})$ certainly is not. The result of section~\ref{sec:derivation} is so useful because this complicated integration is cancelled when using $s$ in maximum-likelihood estimates of $\alpha$ (the cancellation occurs at equation (\ref{eq:cprimeprime})). Furthermore, for FRB searches, the relative fluence $s$ is generally very easy to calculate.

FRB searches regularly report the signal-to-noise ratio, S/N, of events passing a pre-determined threshold, S/N$_{\rm th}$. Calculating both parameters is readily achieved by assuming Gaussian noise statistics for received power in the time--frequency domain, and simply tracking the number of samples added in any given search. In such a case, the parameter $s$ is simply the ratio:
\begin{eqnarray}
s & = & \frac{\rm S/N}{\rm S/N_{\rm th}}, \label{eq:s}
\end{eqnarray}
i.e.\ it is the detected significance relative to threshold significance. Combining equations~(\ref{eq:alpha}) and (\ref{eq:s}) gives:
\begin{eqnarray}
\frac{1}{\alpha^*} & = & \frac{-1}{N} \sum_{i=1}^{N} \log \left( \frac{\rm S/N}{\rm S/N_{\rm th}} \right)_i. \label{eq:alpha_sn}
\end{eqnarray}
It is worthwhile discussing the assumptions in the derivation of section~\ref{sec:derivation} in the context of FRB searches. Certainly, it is possible that the true integral source-counts distribution, $N(F,\bmath{\theta})$, is not separable between `complicating parameters' $\bmath{\theta}$ and the fluence distribution $(F/F_0)^{\alpha}$. An example would be observationally brighter FRBs having lower DMs. However, should this be the case,  the integral source-counts spectrum would almost certainly not be a power-law --- in this specific example, it would necessarily turn over due to the missing bright, high-DM population. Hence, a power-law model would simply be the wrong model to fit, and more-sensitive tests using $s$ will then be more likely to detect deviations from this model.

Assuming that the relative maximum threshold $b$ is constant over all $\bmath{\theta}$ (to derive equation~(\ref{eq:ps})), and set to infinity (to derive equation~(\ref{eq:alpha})), is not a very stringent criterion --- $b$ need only be very large. This is the case for both FRB010724 \citep{2007Sci...318..777L} and FRB180309 \citep{2018ATel11385....1O}, which demonstrate the possibility of an FRB hitting a limit at which its S/N can no longer be estimated in the primary detection beam. In both cases, and likely future cases, it was due to insufficient dynamic range causing saturation in recorded data. For FRB010724, the S/N could be recovered using neighbouring beam detections, while for FRB180309, saturation affected relatively few scintles in the dynamic spectrum. If this effect can be characterised as a moderate linear reduction in an intrinsically high S/N, the resultant effect on $\log s$ for that event would be small. For example, were the true S/N for FRB180309 to be 421 rather than 411, $\log s$ would have increased by 0.01 for that particular event, and hence $\alpha$ increased for an FRB sample by an even smaller fraction. Equivalently, this means that the calculated values of $\alpha$ will be insensitive to any cut-off in the peak FRB luminosity lying far above experimental threshold.

It is also worthwhile noting that the derivation in section~\ref{sec:derivation} (and of \citet{1970ApJ...162..405C}) ignores errors in $s$, i.e.\ it assumes that the measured S/N is unaffected by noise. As discussed in \citet{1973ApJ...183....1M}, when the threshold signal-to-noise ratio S/N$_{\rm th}$ is six or greater, this has negligible effect on the resulting calculations of $\alpha$. The calculation of section~\ref{sec:full} illustrates how to account for these errors.

Some cautionary notes are needed. The value of S/N must be calculated using the detection algorithm which sets S/N$_{\rm th}$, not with a more-detailed follow-up calculation (e.g.\ fitting a detailed pulse profile) that boosts S/N. Doing so would create an artificial gap between the original threshold S/N$_{\rm th}$ and the boosted S/N values, which could not be filled by boosted sub-threshold events, since these are missed by the detection algorithm.

If a data-set is re-examined using a more sensitive \emph{detection} method however, then the newer, updated values of S/N and S/N$_{\rm th}$ \emph{must} be used in the calculation, even for FRBs detected in the prior analyses. The more-sensitive method will only identify new events which were sub-threshold in a previous analyses, equivalent to setting a very low upper sensitivity limit $b$ in equation~(\ref{eq:pre_alpha}). To recover the $b=\infty$ result of equation~(\ref{eq:alpha}) therefore requires using updated values for all events. This applies even if no new events are detected, since changing statistical methods based on observed outcomes (`flip-flopping') leads to biased results.

\subsection{A possible cause of bias}
\label{sec:avoiding}

The most likely cause for bias in calculated values of $s$ however is the use of human `by-eye' discretion to identify plausible candidates against RFI background, as noted by \citet{2018MNRAS.474.1900M} in section~2.3 of that work. When these inspection methods are used to rule out RFI-only candidates, as is usually the case, this does not present a bias. However, when a true candidate is detected in the presence of RFI, then the chance that it is falsely rejected decreases with increasing signal strength.  Mathematically, this implies a time-varying threshold S/N which can be significantly larger than the nominal computed values. If a candidate is accepted this way, then using the nominal S/N$_{\rm th}$ will give an artificially large value of $s$, and hence a biased value of $\alpha$.

Any method to overcome this bias necessarily involves estimating S/N$_{\rm th}$ on an event-by-event basis. The first step requires using a reproducible algorithm (e.g.\ a machine-learning method) to identify FRBs, and associate them with a statistical S/N value. Once an FRB is detected however, it should be fed back into the detection algorithm with artificially reduced S/N, i.e.\ after manually subtracting its fitted flux. This should be repeated until the algorithm no longer identifies the event as an FRB, thus determining the actual S/N$_{\rm th}$ for that event. This method could also work with crowd-sourced classification methods, where the initial and subsequent identifications are performed by a random, and hence independent (but statistically identical) ensemble of volunteers. However, it could not be used when a small number of experts perform by-eye analysis, both because the initial identification will not be replicable, and subsequent identifications of reduced S/N events will not be independent.

\subsection{Equivalence to $V/V_{\rm max}$ test}
\label{sec:vvmax}

The $V/V_{\rm max}$ test of \citet{1968ApJ...151..393S} has been discussed in the context of FRBs by \citet{2016MNRAS.461..984O} and \citet{2018MNRAS.474.1900M}, and applied to a sample of CRAFT FRBs by \citet{craft_nature}. The test calculates for each detection the volume of space over which a source would have been detectable, $V_{\rm max}$, and the volume enclosed within the actual detected distance, $V$. For a non-evolving source distribution, the measured values of $V/V_{\rm max}$ will be uniformly distributed between zero and one, with a mean of $0.5$, and standard deviation of $\frac{1}{2 \sqrt{3}}$ (for a single sample).

In the case of FRBs, where the distance to each source is generally unknown, the value of $V/V_{\rm max}$ for a particular event in Euclidean space can be calculated using:
\begin{eqnarray}
\frac{V}{V_{\rm max}} & = & \frac{R^3}{R_{\rm max}^3} \nonumber \\
\frac{R}{R_{\rm max}} & = & \left(\frac{F}{F_{\rm th}}\right)^{-0.5} \label{eq:vonvmax} \\
\implies \frac{V}{V_{\rm max}} & \equiv & s^{-1.5}.\nonumber
\end{eqnarray}
In other words, the $V/V_{\rm max}$ test uses exactly the same information from each event ($s$) as the calculation of $\alpha$ in equation~(\ref{eq:alpha_sn}). This explains why the $V/V_{\rm max}$ test is insensitive to variations in burst properties and variations in sensitivity, as noted by \citet{craft_nature}.

An interesting exercise is to determine for which kinds of true source distributions each test more effectively rejects the Euclidian, non-evolving hypothesis. In the case of pure power-law distributions, Table~\ref{tab:pwr_law} compares the expected values for each test as a function of $\alpha$, and the rejection power $r$, defined here as:
\begin{eqnarray}
r & = & \lim_{N\to\infty} \sqrt{N} \left( \frac{\mu(\alpha)-\mu(\alpha=-1.5)}{\sigma(\alpha=-1.5)}\right),\label{eq:r}
\end{eqnarray}
where $\mu$ is the expected value of each test, and $\sigma$ is the standard deviation in the case of $\alpha=-1.5$ and number of samples $N$. The limit $N\to\infty$ is taken so that $\sigma$ approaches its large-$N$ form, and biases tend to zero (i.e.\ $\alpha^{\prime} \to \alpha$), while the factor $\sqrt{N}$ re-normalises this to a single sample. From Table~\ref{tab:pwr_law}, the rejection power of the maximum-likelihood test is approximately 30\% greater than that of the $V/V_{\rm max}$ test. This is to be expected, since the maximum-likelihood estimate has been optimised specifically for a power-law distribution.

\begin{table}
\centering
\caption{Rejection power $r$, defined as per equation (\ref{eq:r}), for rejecting the $\alpha=-1.5$ hypothesis, in the case where the true FRB fluence distribution is a power-law with index given by the left-most column. In the $N \to \infty$ limit, this will also equal the expectation value $\mu_l$ for the maximum-likelihood estimation of $\alpha$ from equation (\ref{eq:crawford}). The expectation value of the $V/V_{\rm max}$ test is given by $\mu_v$, while $r_l$ and $r_v$ are the rejection powers of the maximum-likelihood test (equation (\ref{eq:crawford})) and the $V/V_{\rm max}$ test (equation (\ref{eq:vonvmax})) respectively.}\label{tab:pwr_law}
\begin{tabular}{cccc}
\hline
$\alpha (=\mu_l)$ & $\mu_v$ & $r_l$ & $r_v$ \\
\hline
-1.0	& 0.410	& -0.41	&  -0.31 \\
-1.1	& 0.430	& -0.33	& -0.24	\\
-1.2	& 0.449	& -0.25	& -0.18	\\
-1.3	& 0.467	& -0.16	& -0.11	\\
-1.4	& 0.484	& -0.08	& -0.06	\\
-1.5	& 0.500	& 0	& 0	\\
-1.6	& 0.516	& 0.08	& 0.06	\\
-1.7	& 0.531	& 0.16	& 0.11	\\
-1.8	& 0.544	& 0.24	& 0.15	\\
-1.9	& 0.557	& 0.33	& 0.20	\\
-2.0	& 0.570	& 0.41	& 0.24 \\
\hline
\end{tabular}
\end{table}

\section{Calculations with Parkes and ASKAP data}
\label{sec:cpdata}

The methods described above can be applied simultaneously to combined sets of FRB data from multiple telescopes. Here, we limit this to the two instruments with sufficiently large numbers of detected FRBs such that fits to individual telescope data sets will be meaningful. Currently, this is data from various searches with the Parkes radio telescope, and FRBs detected by the Commensal Real-time ASKAP Fast Transients (CRAFT; \citet{2010PASA...27..272M}) survey with ASKAP. In the near future, the upgraded Molonglo Observatory Synthesis Telescope (UTMOST; \citet{2017PASA...34...45B}) and the Canadian Hydrogen Intensity Mapping Experiment (CHIME; \citet{2018ApJ...863...48T}) should both have sufficiently high statistics to apply this method.

Applying equation~(\ref{eq:crawford}) to Parkes and ASKAP data requires only being able to calculate $s$ from equation~(\ref{eq:s}) for each detected FRB, i.e.\ it requires knowing both the detected and instantaneous threshold signal-to-noise ratios.  As discussed in section~\ref{sec:application}, the multitude of potentially complicated effects causing some fraction of FRBs to remain undetected can be ignored.

\begin{figure*}
\begin{center}
\includegraphics[width=\columnwidth]{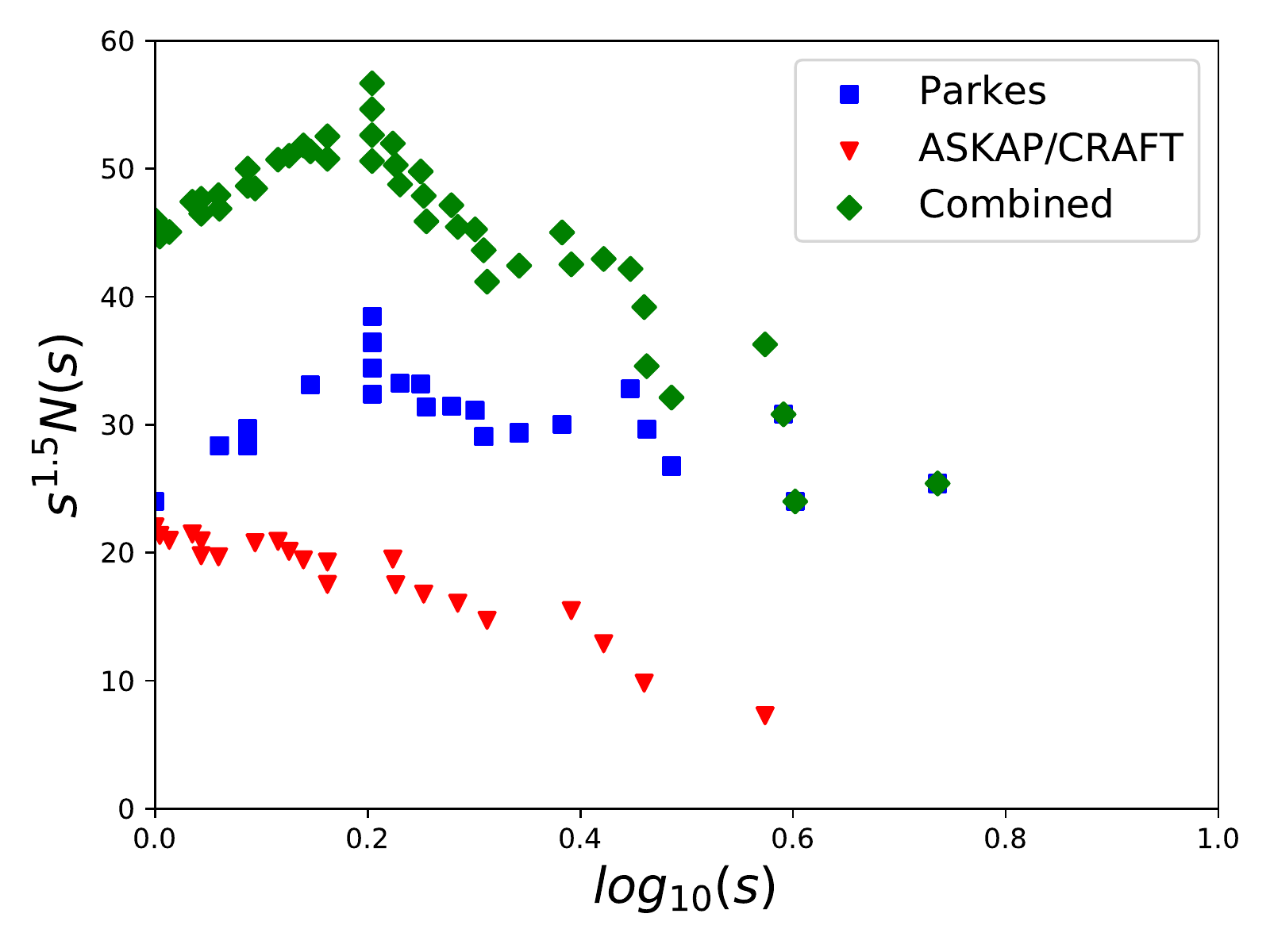}
\caption{
Cumulative source count distributions $N(s)$ of the Parkes, ASKAP/CRAFT, and combined FRB samples used in this work, as a function of the logarithm of relative detection significance $s$. The factor $s^{1.5}$ normalises $N(s)$ to the Euclidean expectation. The point due to FRB 180309 at $\log_{10}(s)=1.61$ is not shown.
} \label{fig:ns}
\end{center}
\end{figure*}

All fast radio bursts detected by Parkes listed on \mbox{FRBCAT} \citep{2016PASA...33...45P} with published ${\rm S/N}_i$ and ${\rm S/N}_{\rm th}$ are given in Table~\ref{tab:snr}. The only exceptions were FRB 150807 \citep{2016Sci...354.1249R}, for which no published S/N or ${\rm S/N}_{\rm th}$ value exists; and the Lorimer Burst (FRB 010724; \citet{2007Sci...318..777L}), which is excluded on grounds of discovery bias \citep{2018MNRAS.474.1900M}.
Equivalent values for ASKAP FRBs detected by CRAFT, which all used ${\rm S/N}_{\rm th}=9.5$ (after post-processing of initial data searched at ${\rm S/N}_{\rm th}=10$), are given in Table~\ref{tab:snrc}. Figure~\ref{fig:ns} shows the distributions of $N(s)$ for these samples.

Using these values as input to equation~(\ref{eq:alpha_sn}), we calculate bias-corrected values $\alpha^{\prime}_p=-1.18 \pm 0.24$ and $\alpha^{\prime}_a=-2.20 \pm 0.47$ for Parkes and ASKAP/CRAFT respectively. The errors are 68\% confidence intervals, calculated using equation~(\ref{eq:gamma_dist}), and scaled according to equation~(\ref{eq:unbiased}) for $\alpha^{\prime}$. They are symmetric at the stated level of precision. Thus the difference corresponds to $1.9$\,$\sigma$ assuming Gaussian errors.

The analysis allows both data-sets to be combined if we assume that the same power-law governs both data-sets. Doing so produces $\alpha^{\prime}_c=-1.55 \pm 0.23$.

\begin{figure*}
\begin{center}
\includegraphics[width=\columnwidth]{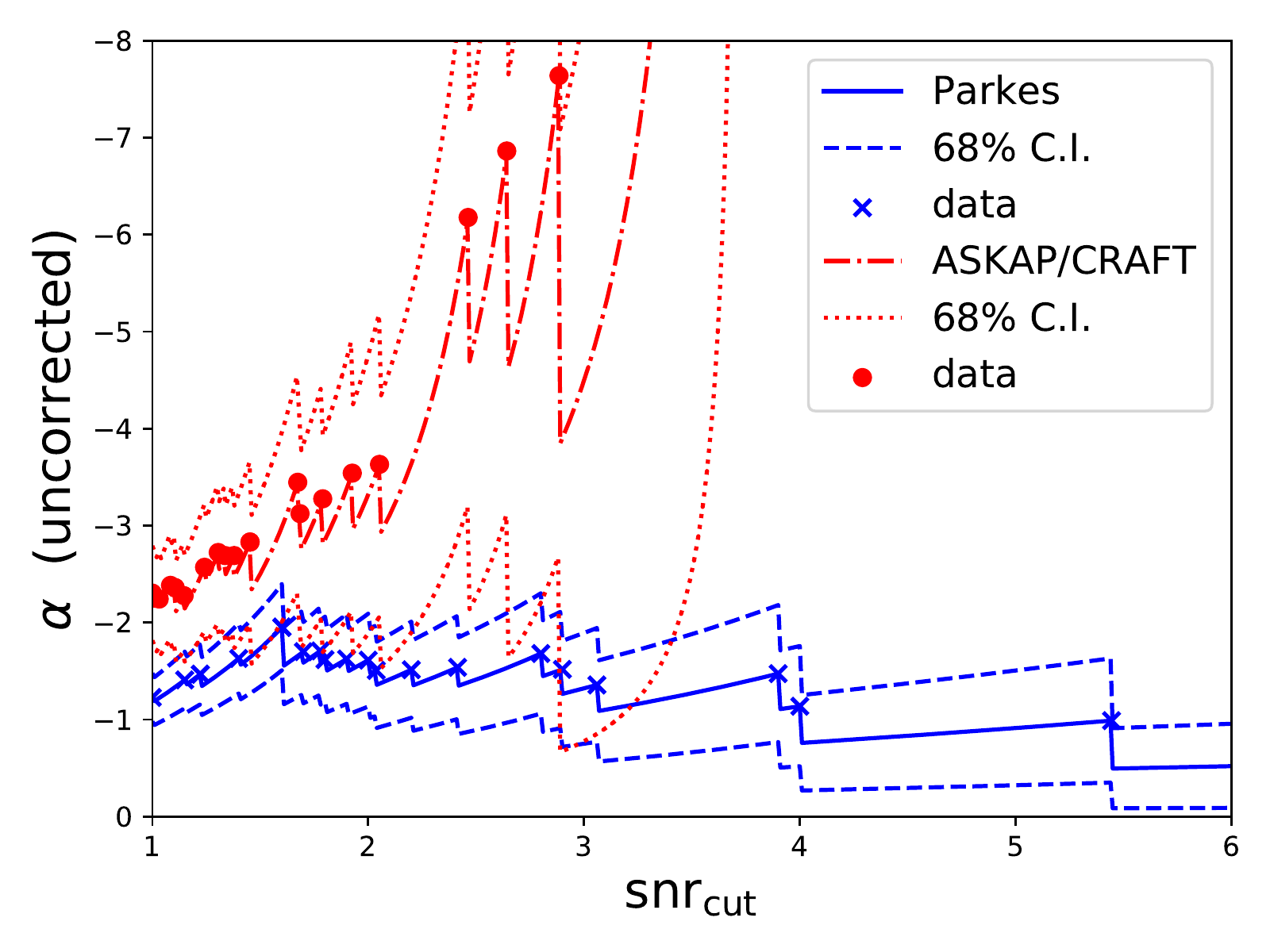} \includegraphics[width=\columnwidth]{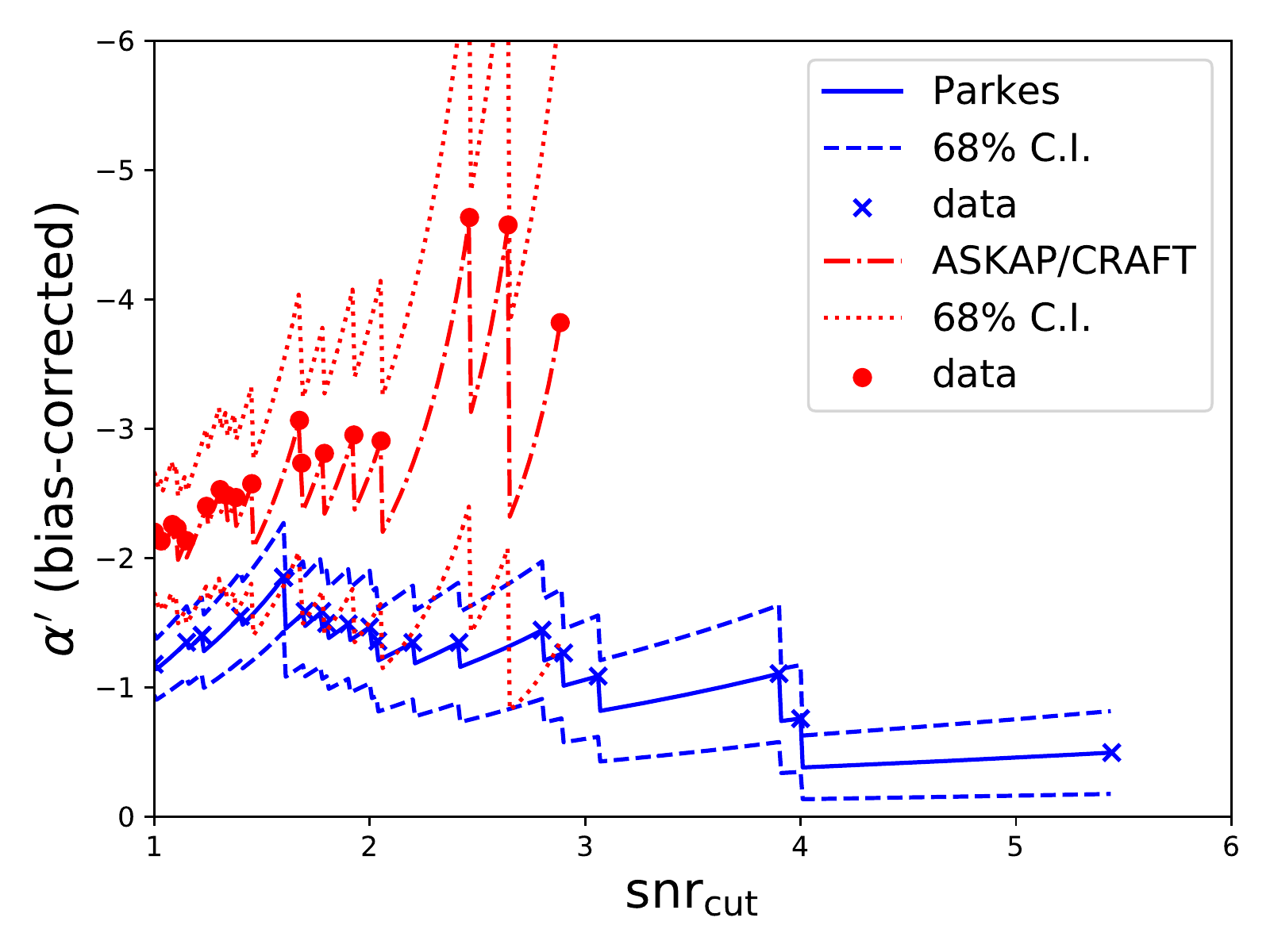}
\includegraphics[width=\columnwidth]{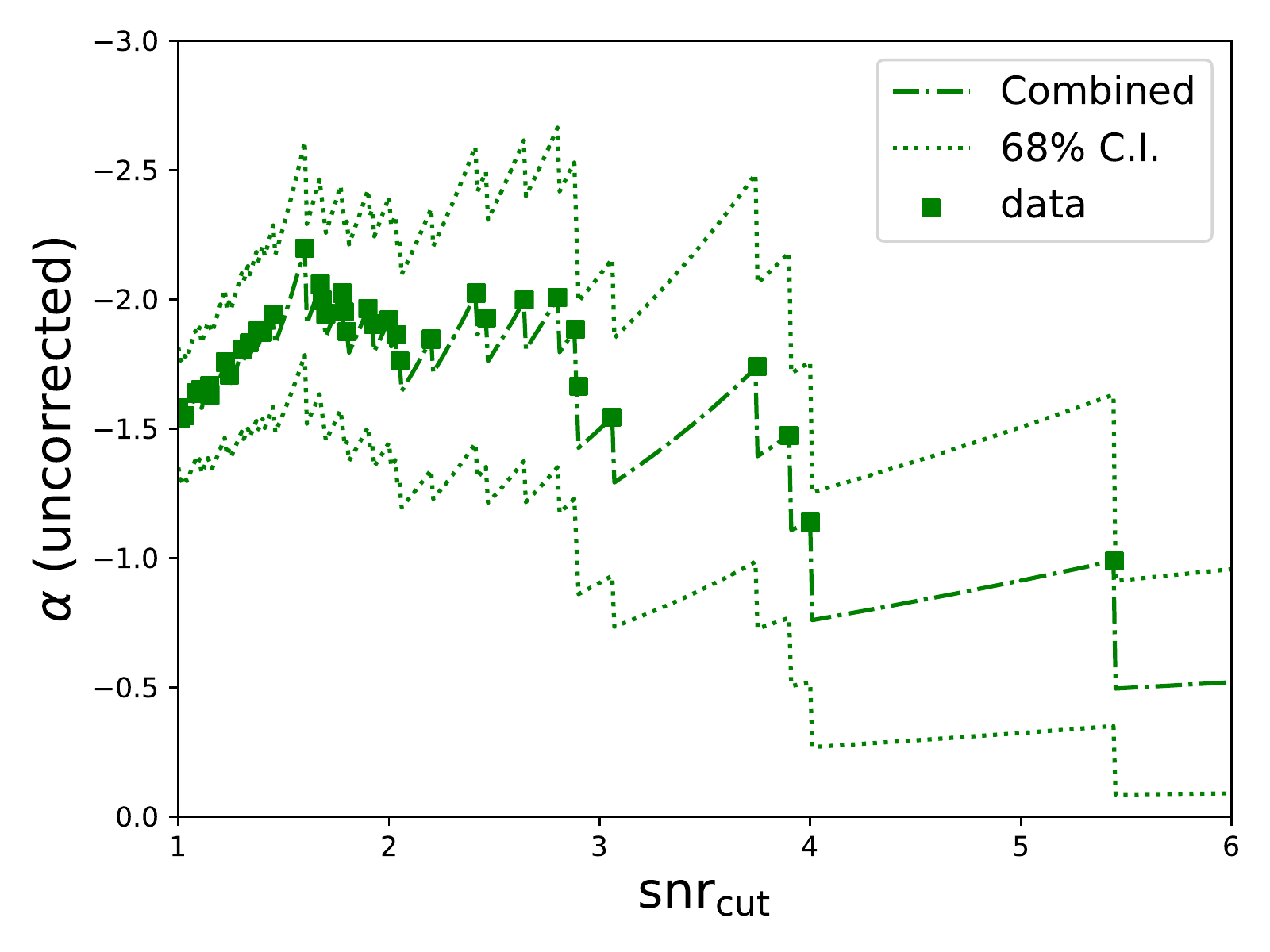} \includegraphics[width=\columnwidth]{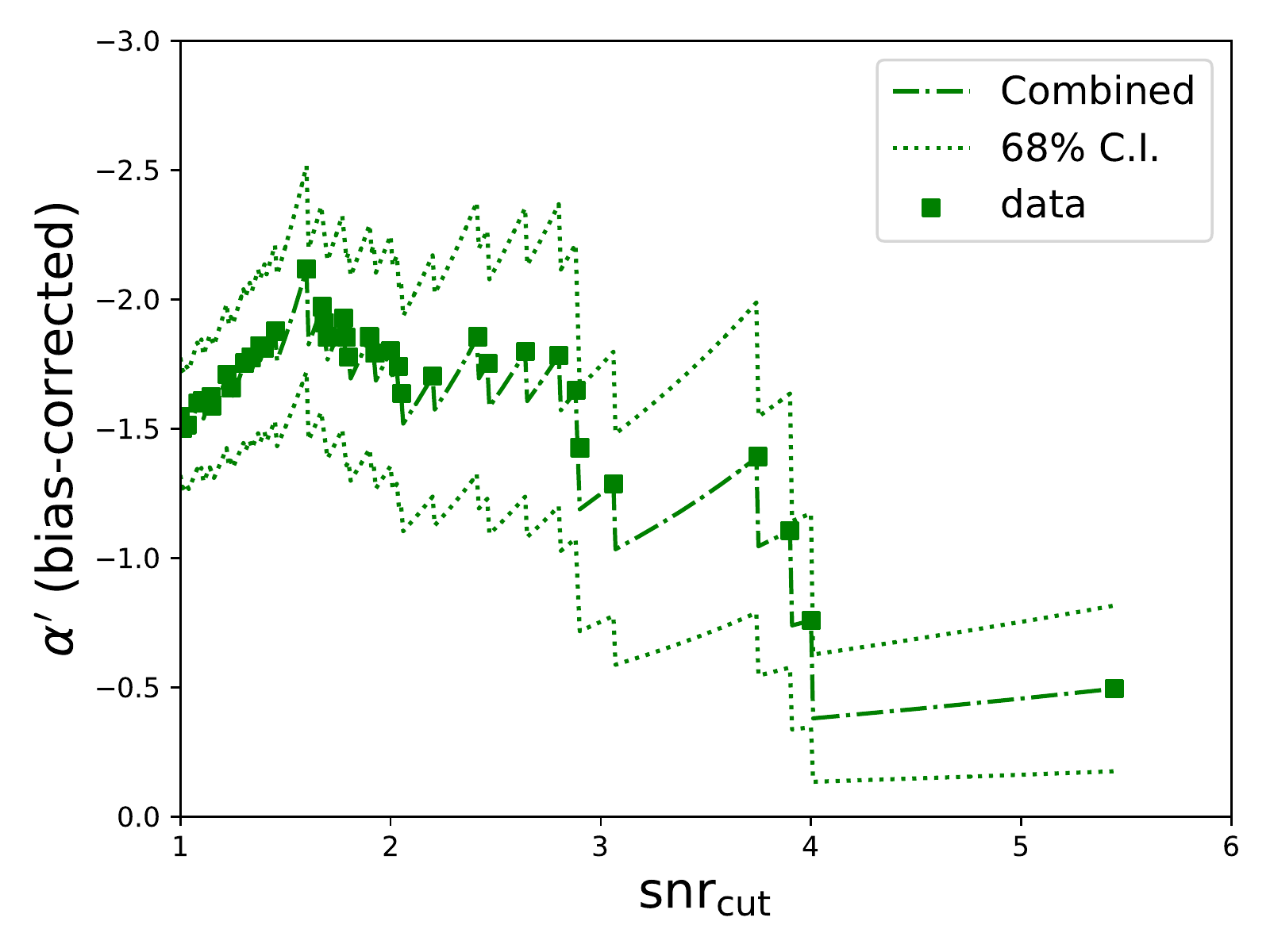} 
\caption{
Calculated values of uncorrected $\alpha$ (left) and corrected $\alpha^{\prime}$ (right), with upper plots using Parkes (blue) and ASKAP/CRAFT (red) data, and lower plots using a combined data set.  Calculations used equation \ref{eq:crawford}, as a function of cutoff signal-to-noise ratio, $s_{\rm cut}$. This was performed as a continuous function of $s_{\rm cut}$ (lines, with error bars according to equation \ref{eq:gamma_dist}), and using values of detected FRBs $s_i$) as the cutoff (points). The uncorrected plots are shown to illustrate the bias on estimates of $\alpha$ when compared to the corrected values.
} \label{fig:analytic_alpha}
\end{center}
\end{figure*}

To check the robustness of the fit, a cutoff in $s$, $s_{\rm cut}$, is introduced. Modified values of $s$, $s_i^{\prime}$, are used in equation~(\ref{eq:crawford}), calculated as:
\begin{eqnarray}
s_i^{\prime} & = & \frac{s_i}{s_{\rm cut}}, \label{eq:sprime}
\end{eqnarray}
with only data points satisfying $s_i^{\prime} \ge 1$ included in the sum. Doing so should produce statistically similar (although correlated) values of $\alpha^{\prime}$, due to the scale-invariance of power-law distributions, while $\alpha$ will slowly vary as per the expected bias. Significantly changing values of $\alpha^{\prime}$ indicate deviations from a power-law, e.g.\ due to near-threshold effects. This procedure is similar to that used by \citet{2018MNRAS.474.1900M} to search for a threshold fluence, where $s_{\rm cut}$ is set to each measured value $s_i$ successively.

Figure~\ref{fig:analytic_alpha} shows the change of both corrected $\alpha^{\prime}$ and uncorrected $\alpha$ as a function of $s_{\rm cut}$. For each data set, changing $s_{\rm cut}$ produces only small variations in $\alpha$ and $\alpha^\prime$, since neighbouring estimates differ by at most one FRB, and are thus highly correlated. This explains why the error ranges are much larger than the variation in the data. While the errors gradually increase, the calculated value for Parkes data, $\alpha^{\prime}_p$, stays approximately constant. There is a small hint of a systematic decrease in $\alpha^{\prime}_p$ for $s_{\rm cut}<1.6$, which is consistent with the notion that RFI is acting to obscure marginal events, as discussed in section~\ref{sec:application}. However, this trend does not appear to be statistically significant, and is followed by a subsequent increase in $\alpha_p$ in the range $1.6<s_{\rm cut}<2$, i.e.\ there is no conclusive evidence for this behaviour in the Parkes data.

This is not the case for ASKAP data however, where the fitted value of $\alpha^{\prime}$ decreases rapidly with the cutoff fluence, i.e.\ the FRB integral source-count distribution appears to be steeper at higher fluences. However, the lower bound of the 68\% C.I.\ stays constant at around $-1.7$, so this also should not be seen as a significant result.

Figure~\ref{fig:analytic_alpha} illustrates the bias that comes with using detected values of $s_i$ as cutoff fluences. Local minima in $\alpha$ are found when $s_{\rm cut}=s_i$, and local maxima when $s_{\rm cut}$ is slightly above $s_i$. Since a point with $s_i=s_{\rm cut}$ contributes nothing to the sum in equation~(\ref{eq:crawford}), but does change the normalisation $N$ by 1, the increase in $\alpha$ when moving from using $N$ to $N-1$ data points is a factor of $(N-1)/N$. Choosing either extremum can thus be associated with a systematic error of $\pm 0.5 \alpha/N$, where the error is negative (positive) for including (excluding) the threshold point $s_i$.

\begin{figure*}
\begin{center}
\includegraphics[width=\columnwidth]{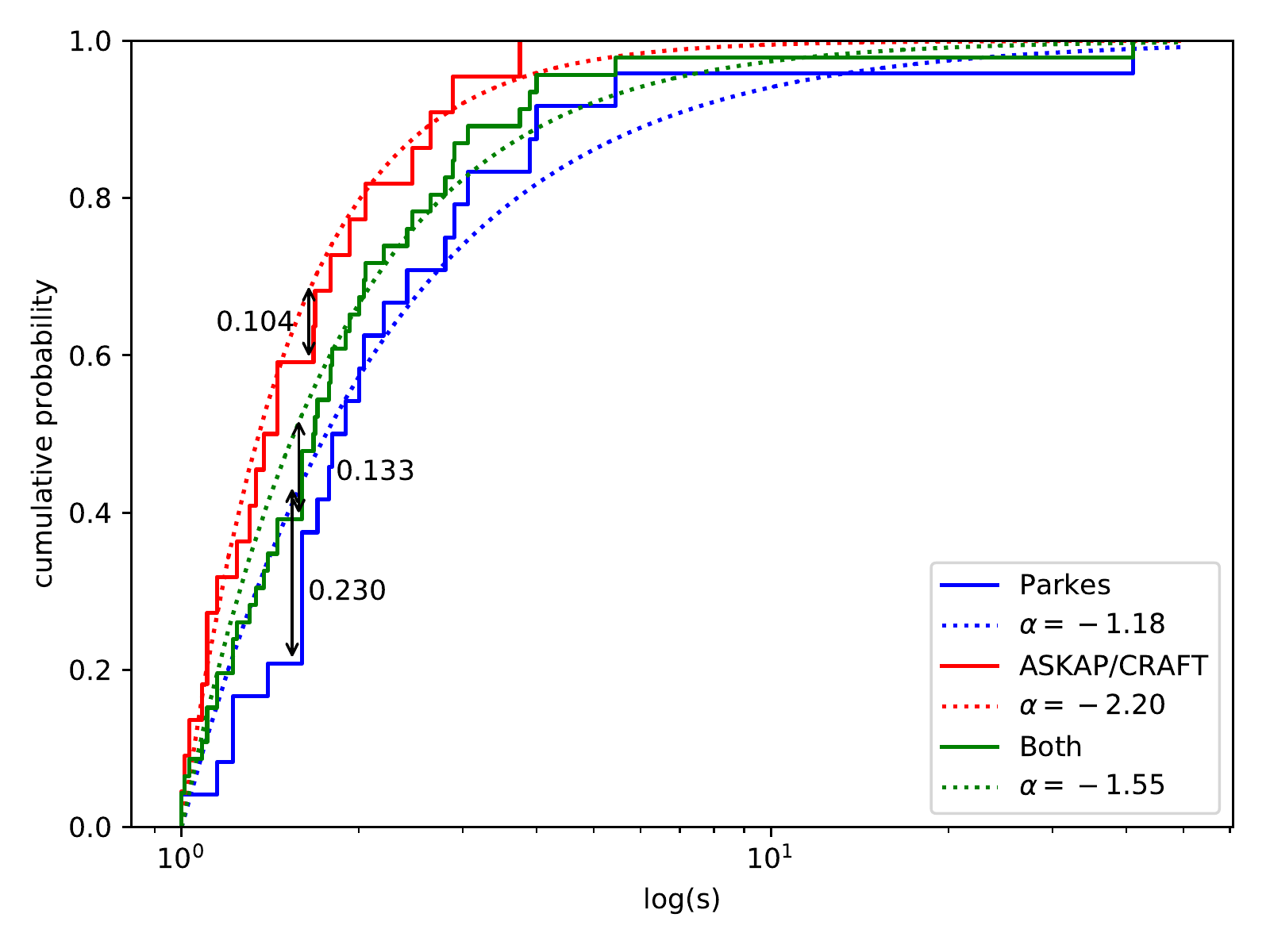} \includegraphics[width=\columnwidth]{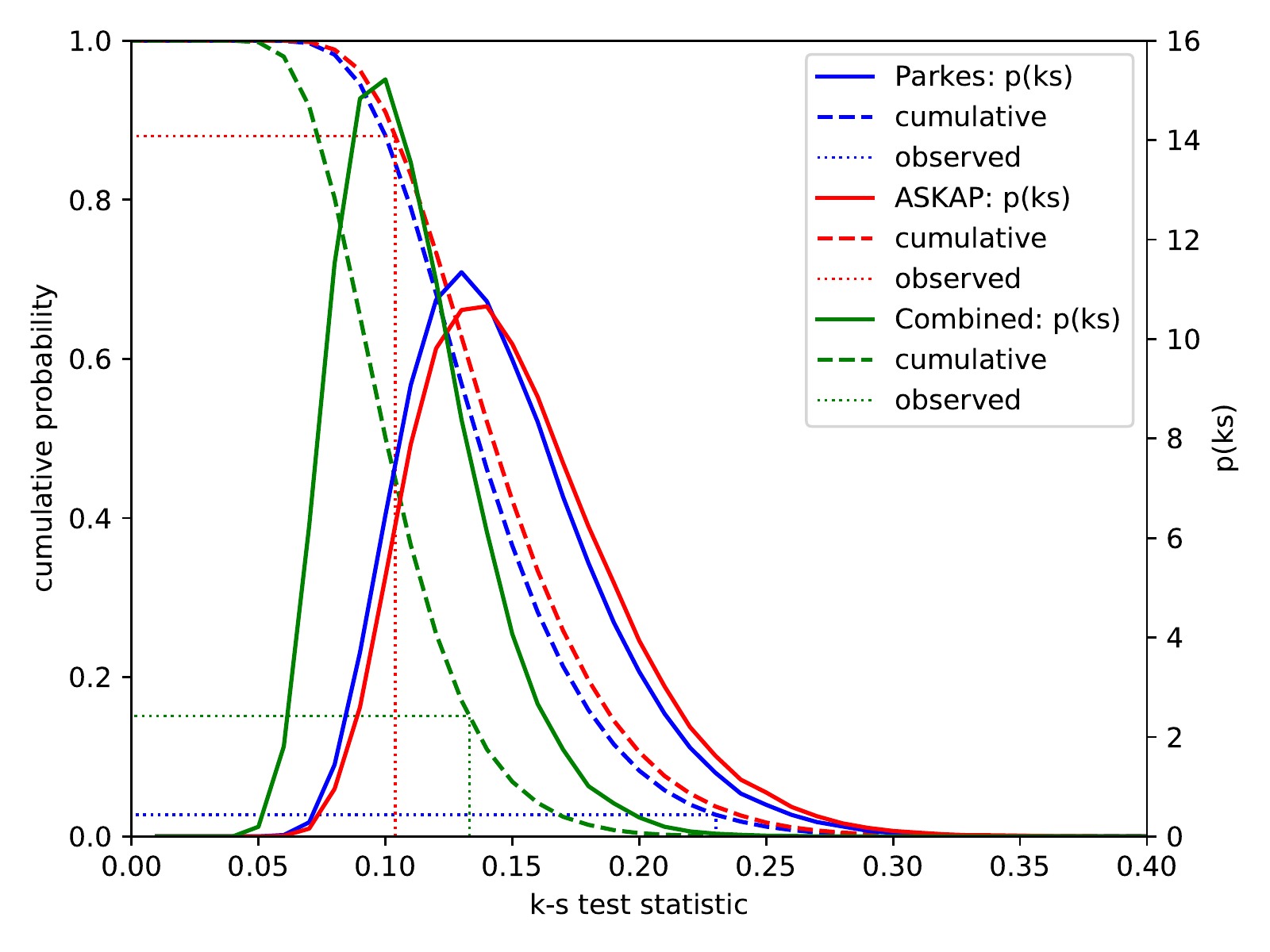}
\caption{Left: Kolmogorov-Smirnov test for compatibility between bias-corrected power-law fits (dotted) to Parkes (blue),  ASKAP (red), and combined (green) data, as a function of the natural log of relative signal-to-noise ratio, $s$ (equation \ref{eq:s}). Maximum deviations, i.e.\ the K--S statistics, are indicated by black arrows. Right: results of a Monte-Carlo estimate of the distributions of the test statistic K--S, assuming true power-laws  equal to the fitted values $\alpha^{\prime}$. Shown are the cumulative (left axis) and differential (right axis) probability distributions for Parkes and ASKAP FRBs, and lines indicating the observed values of the K--S statistic for each.} \label{fig:ks}
\end{center}
\end{figure*}

To test for shape deviations from a pure power law, the Kolmogorov--Smirnov (K--S) test \citep{kolmogorov,smirnov} is performed for each of the data sets, and illustrated in Figure~\ref{fig:ks}. The maximum deviations for each of the Parkes, ASKAP/CRAFT, and combined samples, i.e.\ the K--S statistics, are 0.230, 0.104, and 0.133 respectively.

Since the values of $\alpha$ are estimated from the data, the standard tables of the K--S test statistic under the null hypothesis do not apply. \citet{2004EPJB...41..255G} provide tables of the test statistic for fitted power-law distributions in the case where the estimated value of $\alpha$ is drawn from the data, using the Monte Carlo method. However, their table is sparse, and it is much simpler to perform such a Monte Carlo estimate of the distribution of the K--S statistic under the null hypothesis, which allows a p-value to be assigned. The results of this investigation are given in Figure~\ref{fig:ks} (right). The associated p-values (probability of seeing a K--S statistic equal to or greater than that observed, assuming a true power-law with values given by $\alpha^{\prime}$) are 2.7\% (Parkes), 88\% (ASKAP), and 15\% (combined).\footnote{The trial factor is somewhere between two and three due to the correlated nature of the combined and individual tests, corresponding to between a 5.5\% and 8.2\% chance of any result being as inconsistent as the Parkes data under the null hypothesis of a true power-law distribution.} That is, from the K--S test, there is some evidence that the Parkes sample deviates significantly from an underlying power-law distribution, but none in the case of ASKAP data. The result from the combined case is, as expected, intermediate.

The K--S statistic for Parkes data is a maximum immediately below $s_{\rm cut}=1.6$. Using data at and above $s_{\rm cut}=1.6$ only, we find $\alpha^{\prime}=1.85$, for which there is no significant discrepancy with the value of $\alpha^{\prime}$ found for ASKAP data.

That the K--S statistic highlights the Parkes results as being less consistent with a power-law distribution than the ASKAP/CRAFT data, whereas Figure~\ref{fig:analytic_alpha} suggests the opposite, is likely due to the correlated nature of the data presented in Figure~\ref{fig:analytic_alpha}, and a reminder of the dangers of by-eye interpretations of it.

The K--S test is a very general, and consequently not very powerful, test for differences in the integral source-count spectrum between the two data sets. A full likelihood ratio test is used in section~\ref{sec:full} to quantify this difference. First, we make some comments on previous results.

\subsection{Comments on previous results using completeness fluences}
\label{sec:completeness}
 
The maximum likelihood method of \citet{1970ApJ...162..405C} used in section~\ref{sec:theory} has been invoked by both \citet{2018MNRAS.474.1900M} and \citet{2018MNRAS.475.1427B} on different samples of Parkes FRBs, finding $\alpha=-2.6^{+0.7}_{-1.3}$, and $\alpha=-2.2^{+0.6}_{-1.2}$, for N=9 and 5 FRBs above threshold respectively. In both cases, the values of $s$ inserted into equation~(\ref{eq:crawford}) were the estimated fluences (with no beam correction) relative to a completeness threshold in fluence. \citet{2018MNRAS.474.1900M} calculate $\alpha$ using only observed fluences as thresholds, and hence their estimate of $\alpha$ is biased downwards by $0.5 \alpha/N=0.14$, whereas \citet{2018MNRAS.475.1427B} use a pre-determined experimental completeness threshold of 2\,Jy\,ms. Also correcting for the intrinsic bias in $\alpha$, the adjusted estimates of these authors are $\alpha^{\prime}=-2.2^{+0.55}_{-1}$ and $\alpha^{\prime}=-2.0^{+0.55}_{-1.1}$ respectively (errors are scaled identically to the mean).
 
It is also interesting to comment on the use of a `completeness threshold' $F_0$ with which to normalise observed fluences $F$ in calculating $s$. Firstly, note that in the original historical context (source counts in the $3^{\rm rd}$ Cambridge Catalogue of radio galaxies), it is not readily possible to estimate a signal-to-noise ratio --- see for example the discussion in section~II of \citet{1973ApJ...183....1M}. Thus it was more common to try to find a `completeness flux' (here, fluence) $F_c$ above which all sources in the survey area would have been detected. Applying this notion to the formalism of section~\ref{sec:derivation}, $F_c$ should be set such that:
\begin{eqnarray}
F_c & \ge & F_{\rm th}(\bmath{\theta}) \, \forall \, \bmath{\theta}. \label{eq:fc}
\end{eqnarray}
In this case, $F_c$ can simply replace $F_{\rm th}$ in equation~(\ref{eq:pftheta}), and the basic method of \citet{2018MNRAS.474.1900M} --- increasing $F_c$ until stability is reached in $\alpha$ --- is sound. However, it has several deficiencies in the case of FRB counts.

Firstly, the nominal value of $F_c=2$\,Jy\,ms in the case of most searches at Parkes arises from the \emph{observed} maximum pulse duration of $30$\,ms \citep{2015MNRAS.447.2852K}. However, if $F_{\rm th}$ exceeds the completeness threshold $F_c$ over any part of the parameter space for which FRBs exist (i.e.\ $k(\bmath{\theta})$ is non-zero in this range), then the survey will not in fact be complete. For instance, FRBs of 32\,ms width. The lack of FRBs with observed widths in this range can either be interpreted as validating the $k(\bmath{\theta})=0$ requirement, or as being evidence of incompleteness itself.

Secondly, the criteria of equation~(\ref{eq:fc}) results in many events being rejected, which is unnecessary when their relative signal-to-noise ratio can simply be calculated according to equation~(\ref{eq:s}). In particular, for short FRBs, $2$\,Jy\,ms is well-above the actual detection threshold, which is closer to $0.5$\,Jy\,ms for FRBs of $1$\,ms duration.

Thirdly, equation~(\ref{eq:unbiased}) shows how the bias in $\alpha(F_c)$ increases as the number of events in the sample is reduced. The effect is even greater if the events themselves define the threshold. It induces an artificial slope in plots of $\alpha$ as a function of threshold, which can be seen in Figures~\ref{fig:analytic_alpha}\,(left). Any method which searches for a completeness limit via a flattening in $\alpha(F_c)$ above a critical value will therefore misidentify the completeness limit.

Fourthly, whereas S/N is readily calculable for a FRB search algorithm, care must be taken not to calculate the observed fluence (which is a derived data product) using a more advanced method than that used for the FRB search, e.g.\ by the fitting of and integration over a Gaussian profile. Of course, corrected fluences (e.g.\ beam-corrected values) can yield important information on FRB properties --- they just should not be used for this method.

Finally, when considering the impact of human discretion in identifying candidates against an RFI background, the use of a completeness threshold defined in terms of fluence rather than S/N means that some candidates (e.g.\ those with long durations) with fluence near $F_c$ will indeed be marginal and difficult to identify by eye, while others (e.g.\ those with short durations) will be extremely strong. Artificially increasing $s$ as per equation~(\ref{eq:sprime}) is thus a much more effective way to search for this effect.

The other aspect to note in the method of \citet{2018MNRAS.474.1900M} (and applied by \citet{2018MNRAS.475.1427B}) is that beamshape remains a hidden and uncorrected variable. That is, while a sample may be complete in received fluence, it is not complete in true fluence. Hence, while the former authors note that defining thresholds in terms of S/N leads to incompleteness at a given fluence level, in fact a completeness fluence $F_c$ can not be defined in terms of true FRB fluence either, which will always be greater by some amount than the observed fluence.

In Appendix \ref{sec:extended}, we show --- for completeness --- how this case (defining S/N and/or $F_c$ in terms of detected fluence) reduces to the results of section~\ref{sec:derivation}, and how the threshold method of equation~(\ref{eq:fc}) applies (i.e.\ equally, and with the same caveats). This result can be briefly understood by realising that a detection off beam centre will affect both threshold and measured S/N (or equivalently fluence) by an equal, if unknown, amount, thus preserving their ratio for statistical purposes.

\begin{table}
\centering
\caption{Parkes fast radio bursts with known signal-to-noise ratios S/N, and threshold ratios S/N$_{\rm th}$, allowing relative detection strength $s$ to be calculated.} \label{tab:snr}
\begin{tabular}{l c c c p{3.5cm}}
FRB & S/N & S/N$_{\rm th}$ & s & Ref.\ \\
\hline
110220 & 49 & 9 & 5.44 & \multirow{4}{*}{\citet{2013Sci...341...53T}} \\
110627 & 11 & 9 & 1.22 & \\
110703 & 16 & 9 & 1.78 & \\
120127 & 11 & 9 & 1.22 & \\
\hline
090625 & 28 & 10 & 2.8 & \multirow{5}{*}{\citet{2016MNRAS.460L..30C}} \\
121002 & 16 & 10 & 1.6 & \\
130626 & 20 & 10 & 2 & \\
130628 & 29 & 10 & 2.9 & \\
130729 & 14 & 10 & 1.4 & \\
\hline
150418 & 39 & 10 & 3.9 & \citet{2016Natur.530..453K} \\
\hline
150610 & 18 & 10 & 1.8 & \multirow{4}{*}{\citet{2018MNRAS.475.1427B}} \\
151206 & 10 & 10 & 1 & \\
151230 & 17 & 10 & 1.7 & \\
160102 & 16 & 10 & 1.6 & \\
\hline
010125 & 16.9 & 7 & 2.415 & \citet{2014ApJ...792...19B} \\
\hline
010621 & 16.3 & 8 & 2.04  & \citet{2011MNRAS.415.3065K}; E.~Keane (private communication) \\
\hline
131104 & 30.6 & 10 & 3.06 & \citet{2016Sci...354.1249R} \\
\hline
140514 & 16 & 10 & 1.6 & \citet{2015MNRAS.447..246P} \\
\hline
150215 & 19 & 10 & 1.9 & \citet{2017MNRAS.469.4465P} \\
\hline
171209 & 40 & 10 & 4 & \citet{2017ATel11046....1S} \\
\hline
180301 & 16 & 10 & 1.6 & \citet{2018ATel11376....1P} \\
\hline
180309 & 411 & 10 & 41.1 & \citet{2018ATel11385....1O} \\
\hline
180311 & 11.5 & 10 & 1.15 & \citet{2018ATel11396....1O} \\
\hline 
180714 & 22 & 10 & 2.2 & \citet{2018ATel11851....1O} \\
\hline
\end{tabular}
\end{table}

\begin{table}
\centering
\caption{CRAFT FRBs, showing detected signal-to-noise ratios S/N, and relative ratio $s$, compared to the detection threshold S/N$_{\rm th}=9.5$.} \label{tab:snrc}
\begin{tabular}{c c c}
\hline
FRB & S/N  & $s$ \\
\hline
\multicolumn{3}{c}{\citet{2017ApJ...841L..12B}} \\
170107 & 16 & 1.68 \\
\hline
\multicolumn{3}{c}{\citet{craft_nature}} \\
\hline
170416 & 13.1 & 1.38 \\
170428 & 10.5 & 1.11 \\
170712 & 12.7 & 1.34 \\
170707 & 9.5 & 1.00 \\
170906 & 17 & 1.79 \\
171003 & 13.8 & 1.45 \\
171004 & 10.9 & 1.15 \\
171019 & 23.4 & 2.46 \\
171020 & 11. & 2.05 \\
171116 & 11.8 & 1.24 \\
171213 & 25.1 & 2.64 \\
180110 & 35.6 & 3.75 \\
180119 & 15.9 & 1.67 \\
180128.0 & 12.4 & 1.31 \\
180128.2 & 9.6 & 1.01 \\
180130 & 10.3 & 1.08 \\
180131 & 13.8 & 1.45 \\
180212 & 18.3 & 1.93 \\
\hline
\multicolumn{3}{c}{\citet{jpnew}} \\
\hline
180315 & 10.5 & 1.11 \\
180324 & 9.8 & 1.03 \\
180525 & 27.4 & 2.88 \\
\hline
\end{tabular}
\end{table}

\section{Full maximum-likelihood calculation of Parkes and ASKAP FRBs in the presence of noise}
\label{sec:full}

We now present a traditional maximum-likelihood estimate of $\alpha$, and use a likelihood ratio test to determine whether separate power laws describe ASKAP/CRAFT and Parkes data significantly better than a single power-law.

The methods used in section~\ref{sec:theory} ignore the contribution of noise to the measured value of signal-to-noise ratio S/N, and hence $s$. Such a case is covered by \citet{1973ApJ...183....1M}, where the authors note that when the threshold S/N is $6$ or greater, no significant effect due to noise is expected when calculating the power-law index $\alpha$. Here, we explicitly check this using a full likelihood maximisation.

The likelihood of observing a signal-to-noise ratio $S_{\rm obs}$ for a true power-law distribution of $S_{\rm true}$ depends on the noise deviate $n=S_{\rm obs}-S_{\rm true}$:
\begin{eqnarray}
P(S_{\rm obs}) & = & \frac{1}{C} \int_{-\infty}^{+\infty} P(n)\,dn \left( \frac{S_{\rm true}}{S_{\rm th}} \right)^{\alpha}, \label{p_sobs}
\end{eqnarray}
where the normalisation constant $C$ is the probability of observing any such event above a threshold $S_{\rm th}$:
\begin{eqnarray}
C & = & \int_{S_{\rm th}}^{\infty} P(S_{\rm obs}) \, d S_{\rm obs}.
\end{eqnarray}
For standard normal deviates, $P(S_{\rm obs})$ is simply:
\begin{eqnarray}
P(S_{\rm obs}) & = & \frac{1}{C} \int_{-\infty}^{+\infty} \frac{1}{\sqrt{2 \pi}} e^{-( 0.5 n^2)} dn \left( \frac{S_{\rm true}}{S_{\rm th}} \right)^{\alpha}. \label{eq:plike}
\end{eqnarray}
The integral over $dn$ can in practice be limited to a small range --- here, $\pm 5$ is used. This has been tested to reproduce the values of \citet{1973ApJ...183....1M} (Table~1) to within the stated precision of four significant figures.

The simplest definition of the likelihood function $\mathcal{L}$ for the Parkes and ASKAP samples of FRBs is therefore:
\begin{eqnarray}
\mathcal{L} & = & \frac{1}{N} \sum_{i=1}^{N} \log P(S_{\rm obs}^i), \label{eq:ell}
\end{eqnarray}
where the sum proceeds over all $i$ observations. The $P_i$ are calculated according to equation~(\ref{eq:plike}), with $S_{\rm obs}$ and $S_{\rm th}$ the measured and threshold signal-to-noise ratios for that particular observation (the constant $C$ must also be re-normalised for each observation). This is only applicable when using Parkes data; the ASKAP data used a constant threshold S/N of $9.5\sigma$.

A more-correct estimate in the case of Parkes or combined data would require weighting each observation by the relative fraction of observation time spent observing at that given threshold. However, these values are generally not available, and the expected loss of precision for a small variation in $S_{\rm th}$ is also small.

\begin{figure}
\centering
\includegraphics[width=\columnwidth]{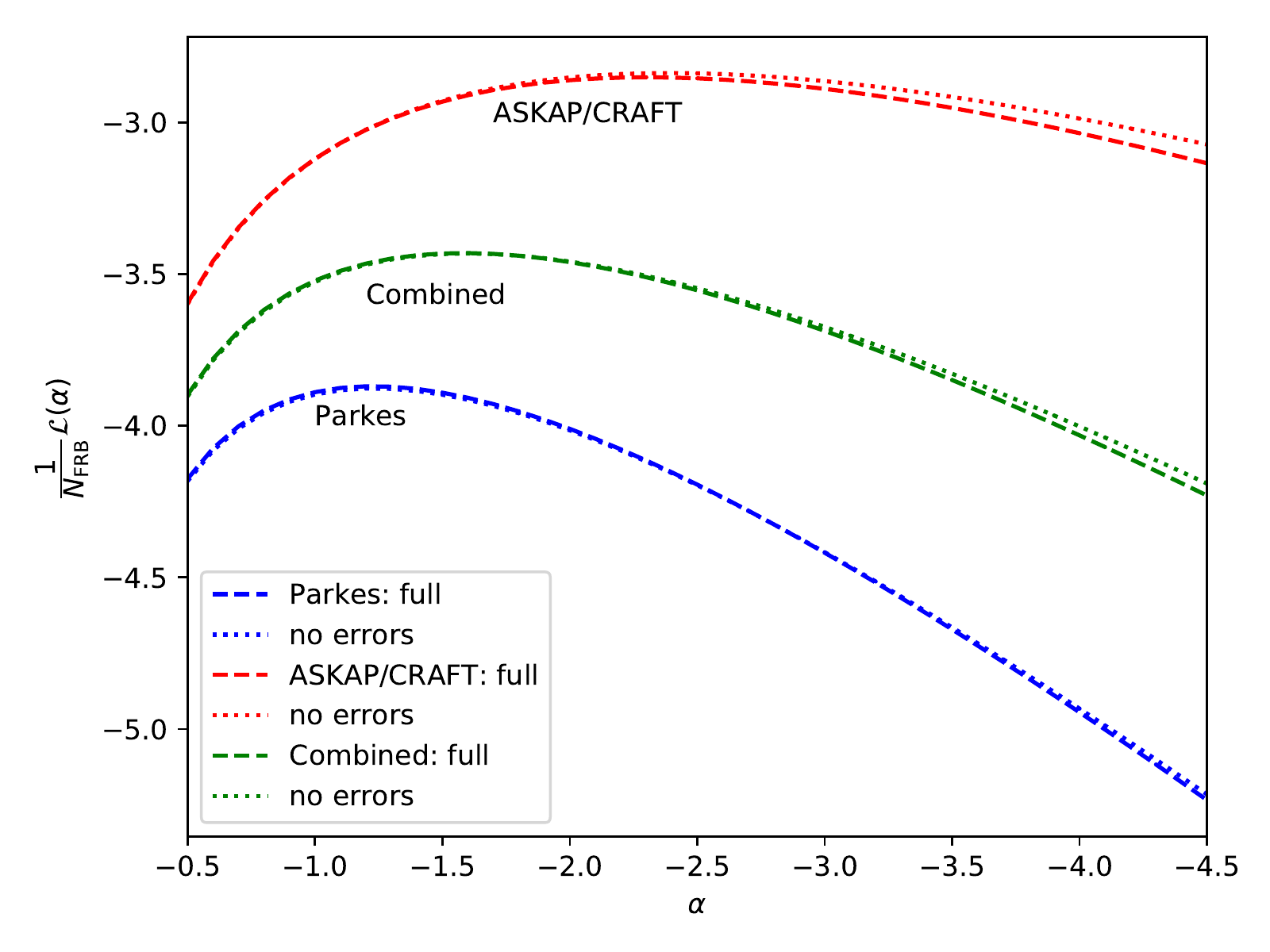}
\caption{Evolution of the maximum likelihood $\mathcal{L}$ as a function of $\alpha$, for Parkes, ASKAP/CRAFT, and combined data sets. Dashed lines use the full Gaussian error distribution from equation~(\ref{eq:plike}), while dotted lines ignore the noise contribution to $S_{\rm obs}$.}\label{fig:max_ell}
\end{figure}

Calculating $\mathcal{L}$ as a function of $\alpha$ for the ASKAP, Parkes, and combined samples (Tables~\ref{tab:snr} and \ref{tab:snrc}) produces the values shown in Figure~\ref{fig:max_ell}. To estimate the influence of Gaussian noise, calculations were also performed by setting $S_{\rm obs} = S_{\rm true}$, i.e.\ $P(S_{\rm obs}) = \delta(S_{\rm obs}-S_{\rm true})$. In general, this was found to have negligible effect on the resulting likelihoods, except for large values of $\alpha$.

The best fit values of $\alpha$ are found at the the maximum of $\mathcal{L}$, $\mathcal{L}^{\rm max}$. For the Parkes, ASKAP/CRAFT, and combined data-sets, these were $-1.24$, $-2.40$, and $-1.61$ respectively. These are consistent with both corrected and uncorrected values calculated in section~\ref{sec:application}, although the uncorrected values are the proper comparators (the bias of the analytic method is inherent in the maximum-likelihood procedure).

While this numerical method does not allow confidence intervals to be set for $\alpha$, it does allow the likelihood ratio test to be performed. This is because the combined fit represents a constricted model of individual fits to ASKAP and Parkes data. The test statistic $\mathcal{D}$ is defined as:
\begin{eqnarray}
\mathcal{D} & = & -2 \log \left( \frac{\mathcal{L}_{\rm c}^{\rm max} }{ \mathcal{L}_{a}^{\rm max} + \mathcal{L}_{p}^{\rm max} } \right),
\end{eqnarray}
where `p', `a', and `c' denote Parkes, ASKAP, and combined data respectively. According to Wilks' theorem \citep{wilks1938}, as the number of observations tend to infinity, the distribution of $\mathcal{D}$ under the null hypothesis (that both Parkes and ASKAP data come from the same power-law distribution) will approach a $\chi^2_1$ distribution. The number of degrees of freedom is one due to the single extra power-law being fitted.

For this case, $\mathcal{D}=6.9$, corresponding to a one-sided p-value of 0.86\% for a $\chi^2_1$ distribution. That is, there is evidence at the $2.6\,\sigma$ level that the Parkes and ASKAP data originate from distributions with power-laws of different indices, rather than the same index.

Note that this test is complementary to the K--S test presented in section~\ref{sec:application}, which tests for deviations from power-law-like behaviour. In the specific case of testing one vs.\ two power laws however, this test is more sensitive, since it tests a specific deviation from single power-law behaviour, rather than the general case. Hence, greater inconsistency with the null hypothesis of both Parkes and ASKAP data coming from the same power-law distribution is found.

Specifically testing for differences of each sample with a Euclidean distribution ($\alpha=-1.5$), the test statistic becomes:
\begin{eqnarray}
\mathcal{D} & = & -2 \log \left( \frac{\mathcal{L}_{\rm p/c}(\alpha=-1.5)}{ \mathcal{L}_{\rm p/c}^{\rm max}} \right).
\end{eqnarray}
This produces $D=1.01$ for Parkes data (no evidence for deviations from a Euclidean distribution) and $D=4.00$ for ASKAP/CRAFT data, corresponding to a one-sided 5\% probability of a more extreme deviation in the case of a Euclidean distribution.

\section{DISCUSSION}
\label{sec:discussion}

\subsection{Evidence of a break in the FRB source-count distribution}

Using the likelihood-ratio test, we find evidence (p-value $0.86$\%, i.e.\ 2.6\,$\sigma$) that the Parkes and ASKAP/CRAFT samples arise from distributions with different power-law indices, of $-1.22 \pm 0.25$ and $-2.15 \pm 0.49$ respectively (68\% C.I.). Our ASKAP/CRAFT result agrees with the value of $-2.1^{+0.6}_{-0.5}$ found by \citet{craft_nature} using an almost identical sample, but with the less-sensitive $V/V_{\rm max}$ test.

The nominal Parkes threshold for FRB detection is $0.5$\,Jy\,ms for a 1\,ms duration FRB \citep{2018MNRAS.473..116K}, while for ASKAP, it is 26\,Jy\,ms \citep{craft_nature}. Therefore, if the FRB source count distribution observed by ASKAP is steeper, Parkes cannot observe a pure power-law distribution. This is confirmed by our K--S test results, which show that only 2.7\% of true power-law distributions would show a worse fit to Parkes data.

Under the null hypothesis of Parkes and ASKAP/CRAFT data coming from the same power-law distribution, both p-values --- denoted as $p_1=0.86\%$ and $p_2=2.7\%$ --- will be independent and uniformly distributed (this has been confirmed with simulations). Their product $p_1 p_2$ will therefore have a cumulative distribution of $p_1 p_1 (1- \ln p_1 p_2)$, with low values providing evidence for the alternative hypothesis of deviation from power-law behaviour. This one-sided test provides 0.22\% ($3.1 \sigma$) evidence against the null hypothesis. We have performed this calculation \emph{a posteriori} however, and therefore consider our \emph{a priori} probability of $0.86$\% more reliable.

The K--S test of the ASKAP/CRAFT S/N distribution tests for the steepening occurring above the ASKAP detection threshold. Since we find this data to be well-fit by a power-law (p-value $86\%$), this is consistent with the steepening occurring over the fluence range between the Parkes and ASKAP/CRAFT thresholds.

Another interpretation of our results is the potential bias discussed in section~\ref{sec:avoiding}, whereby effective experimental thresholds are significantly larger than the claimed thresholds. If this is correct, then Parkes data below $s=1.6$ should be excluded, i.e.\ the true effective experimental threshold for most Parkes observations must be near S/N=$16$, given for the most common Parkes threshold of $10\,\sigma$.

Further evidence for a turn-over in the observed Parkes rates however can be found in the literature. The values of $\alpha$ found for Parkes data by both \citet{2018MNRAS.474.1900M} and \citet{2018MNRAS.475.1427B}, after correcting for biases as discussed in section~\ref{sec:completeness}, are $\alpha=-2.2^{+0.55}_{-1}$ and $\alpha=-2.0^{+0.55}_{-1.1}$ respectively. While these estimates use an artificially high fluence threshold due to the use of a completeness limit, the calculations --- after bias correction --- are still statistically valid. Their results therefore appear to be in tension with our value of $\alpha=-1.22 \pm 0.25$ for Parkes.
However, the higher fluence threshold samples a fluence regime more similar to that seen by ASKAP, so that if a break in the source-count spectrum is present, then the used completeness threshold likely lies above it. This then explains the better agreement with the ASKAP value of $\alpha$ found here despite the use of data from Parkes.

The effective threshold at $\alpha=-2.2$ of ASKAP/CRAFT observations is 40\,Jy\,ms to a $1.2656$\,ms pulse \citep{ME}. For Parkes, the nominal threshold of 0.5\,Jy\,ms would be increased to an effective threshold of at least 5\,Jy\,ms for $\alpha > -1.5$ and a Gaussian beamshape \citep{ME}. Therefore, we expect the downturn in the FRB fluence distribution, $F_b$, to lie in this range.

\subsection{Interpretation}

In this subsection, we consider what our results would imply about the nature of FRBs, should they be verified by further data.

\citet{2018MNRAS.tmp.1976M} discuss how features in the FRB source-count distribution will be intimately related to the cosmological evolution of the FRB progenitors. A value of $\alpha=-2.15$ found here is typical of a flat FRB spectral index $s_\nu$ ($F \propto \nu^{s_\nu}$ for frequency $\nu$) and source evolution function which is more strongly peaked than the star-formation rate. It is analogous to the high-fluence limit of AGN source counts (e.g.\ \citet{1980MNRAS.193..683W}).

\citet{2018MNRAS.tmp.1976M} also note that a flat distribution in FRB energy/luminosity up to some maximum $E_{\rm max}/L_{\rm max}$, combined with a downturn in cosmological source evolution (at redshift $z_0$), will lead to a break in the observed differential source-count distribution. The break is predicted to be at a fluence $F_b = E_{\rm max} (1+z_0)^{2+s_\nu}(4 \pi D_L^2(z_0)^{-1})$, where $D_L$ is the luminosity distance.

FRB data is currently too sparse to produce a full fit of the FRB luminosity function. However, it is useful to check whether existing constraints on $F_b$, $E_{\rm max}$, $z_0$, and $s_\nu$ can be mutually consistent with the model of \citet{2018MNRAS.tmp.1976M}. \citet{craft_nature} observe that ASKAP/CRAFT FRBs are viewed out to approximately $z \sim 1$, and Parkes FRBs out to $z \sim 2$--$3$ (suggesting $1 \le z_0 \le 3$), and that the most energetic FRBs detected by ASKAP/CRAFT, Parkes, and UTMOST have energies $E_{\rm max}$ around $10^{33}$-$10^{34}$\,erg\,Hz$^{-1}$. \citet{jpnew} find $s_\nu = -1.5 \pm 0.3$ for the ASKAP/CRAFT sample. These ranges for $s_\nu$, $E_{\rm max}$, and $z_0$ are indeed generally consistent with $5 \le F_b \le 40$\,Jy\,ms, although some regions of the parameter space can be excluded. The observed ranges are not hard limits however, and further analyses and/or observations will be required to further constrain these parameters.

\section{Conclusions}

We have extended the statistical results of \citet{1970ApJ...162..405C}, allowing surveys with unknown completeness thresholds in physical units such as flux/fluence, but readily definable detection and threshold values of signal-to-noise, to calculate the slope of power-law distributions of detected events.

Applied to fast radio bursts (FRBs), this allows improved estimates of the slope of the integral source-count distribution, $\alpha$, without the need to consider a completeness limit, beam pattern, or any other confounding factor. Combining detections with the Parkes radio telescope and CRAFT detections using ASKAP, we find a bias-corrected value for $\alpha$ of $-1.52 \pm 0.24$. Fitting these samples individually however, we find $-1.18 \pm 0.24$ and $-2.20 \pm 0.47$, respectively. A likelihood-ratio test indicates this is compatible with a single power-law at a p-value of $0.86$\%, and that the ASKAP/CRAFT sample is only compatible with the Euclidean expectation of $-1.5$ at a 5\% level. This implies either a steepening in the FRB luminosity function in the fluence range 5--40\,Jy\,ms, or that the experimental threshold for most Parkes FRBs is significantly higher than reported, being approximately $16\,\sigma$. This is the first hint of structure in the FRB source-count distribution, which could be the first evidence of a turn-over in a sharply peaked source evolution of FRB progenitors in the redshift range 1--3.

\section*{Acknowledgements}

We thank S.~Bhandari for discussion on the analysis of Parkes FRBs detected by SUPERB. R.M.S.\ acknowledges support through Australian Research Council (ARC) grants FL150100148 and CE170100004. This work was supported by resources provided by The Pawsey Supercomputing Centre with funding from the Australian Government and the Government of Western Australia. Parts of this research  were conducted by the Australian Research Council Centres of Excellence for All Sky Astrophysics (CAASTRO, CE110001020). This research was also supported by the Australian Research Council through grant DP18010085.

\bibliographystyle{mnras}
\bibliography{bibtex_entries.bib}

\newcommand{\noop}[1]{}
\begin{thebibliography}{}
\makeatletter
\relax
\def\mn@urlcharsother{\let\do\@makeother \do\$\do\&\do\#\do\^\do\_\do\%\do\~}
\def\mn@doi{\begingroup\mn@urlcharsother \@ifnextchar [ {\mn@doi@}
  {\mn@doi@[]}}
\def\mn@doi@[#1]#2{\def\@tempa{#1}\ifx\@tempa\@empty \href
  {http://dx.doi.org/#2} {doi:#2}\else \href {http://dx.doi.org/#2} {#1}\fi
  \endgroup}
\def\mn@eprint#1#2{\mn@eprint@#1:#2::\@nil}
\def\mn@eprint@arXiv#1{\href {http://arxiv.org/abs/#1} {{\tt arXiv:#1}}}
\def\mn@eprint@dblp#1{\href {http://dblp.uni-trier.de/rec/bibtex/#1.xml}
  {dblp:#1}}
\def\mn@eprint@#1:#2:#3:#4\@nil{\def\@tempa {#1}\def\@tempb {#2}\def\@tempc
  {#3}\ifx \@tempc \@empty \let \@tempc \@tempb \let \@tempb \@tempa \fi \ifx
  \@tempb \@empty \def\@tempb {arXiv}\fi \@ifundefined
  {mn@eprint@\@tempb}{\@tempb:\@tempc}{\expandafter \expandafter \csname
  mn@eprint@\@tempb\endcsname \expandafter{\@tempc}}}

\bibitem[\protect\citeauthoryear{{Bailes} et~al.,}{{Bailes}
  et~al.}{2017}]{2017PASA...34...45B}
{Bailes} M.,  et~al., 2017, \mn@doi [\pasa] {10.1017/pasa.2017.39}, \href
  {http://adsabs.harvard.edu/abs/2017PASA...34...45B} {34, e045}

\bibitem[\protect\citeauthoryear{{Bannister} et~al.,}{{Bannister}
  et~al.}{2017}]{2017ApJ...841L..12B}
{Bannister} K.~W.,  et~al., 2017, \mn@doi [\apjl] {10.3847/2041-8213/aa71ff},
  \href {http://adsabs.harvard.edu/abs/2017ApJ...841L..12B} {841, L12}

\bibitem[\protect\citeauthoryear{{Bhandari} et~al.,}{{Bhandari}
  et~al.}{2018}]{2018MNRAS.475.1427B}
{Bhandari} S.,  et~al., 2018, \mn@doi [\mnras] {10.1093/mnras/stx3074}, \href
  {http://adsabs.harvard.edu/abs/2018MNRAS.475.1427B} {475, 1427}

\bibitem[\protect\citeauthoryear{{Burke-Spolaor} \&
  {Bannister}}{{Burke-Spolaor} \& {Bannister}}{2014}]{2014ApJ...792...19B}
{Burke-Spolaor} S.,  {Bannister} K.~W.,  2014, \mn@doi [\apj]
  {10.1088/0004-637X/792/1/19}, \href
  {http://adsabs.harvard.edu/abs/2014ApJ...792...19B} {792, 19}

\bibitem[\protect\citeauthoryear{{Champion} et~al.,}{{Champion}
  et~al.}{2016}]{2016MNRAS.460L..30C}
{Champion} D.~J.,  et~al., 2016, \mn@doi [\mnras] {10.1093/mnrasl/slw069},
  \href {http://adsabs.harvard.edu/abs/2016MNRAS.460L..30C} {460, L30}

\bibitem[\protect\citeauthoryear{{Chatterjee} et~al.,}{{Chatterjee}
  et~al.}{2017}]{2017Natur.541...58C}
{Chatterjee} S.,  et~al., 2017, \mn@doi [\nat] {10.1038/nature20797}, \href
  {http://adsabs.harvard.edu/abs/2017Natur.541...58C} {541, 58}

\bibitem[\protect\citeauthoryear{{Crawford}, {Jauncey}  \&
  {Murdoch}}{{Crawford} et~al.}{1970}]{1970ApJ...162..405C}
{Crawford} D.~F.,  {Jauncey} D.~L.,   {Murdoch} H.~S.,  1970, \mn@doi [\apj]
  {10.1086/150672}, \href {http://adsabs.harvard.edu/abs/1970ApJ...162..405C}
  {162, 405}

\bibitem[\protect\citeauthoryear{{Goldstein}, {Morris}  \& {Yen}}{{Goldstein}
  et~al.}{2004}]{2004EPJB...41..255G}
{Goldstein} M.~L.,  {Morris} S.~A.,   {Yen} G.~G.,  2004, \mn@doi [European
  Physical Journal B] {10.1140/epjb/e2004-00316-5}, \href
  {http://adsabs.harvard.edu/abs/2004EPJB...41..255G} {41, 255}

\bibitem[\protect\citeauthoryear{{James} et~al.,}{{James} et~al.}{2018}]{ME}
{James} C.,  et~al., \noop{3001}Submitted to PASA, 2018, \pasa

\bibitem[\protect\citeauthoryear{{Keane} \& {Petroff}}{{Keane} \&
  {Petroff}}{2015}]{2015MNRAS.447.2852K}
{Keane} E.~F.,  {Petroff} E.,  2015, \mn@doi [\mnras] {10.1093/mnras/stu2650},
  \href {http://adsabs.harvard.edu/abs/2015MNRAS.447.2852K} {447, 2852}

\bibitem[\protect\citeauthoryear{{Keane}, {Kramer}, {Lyne}, {Stappers}  \&
  {McLaughlin}}{{Keane} et~al.}{2011}]{2011MNRAS.415.3065K}
{Keane} E.~F.,  {Kramer} M.,  {Lyne} A.~G.,  {Stappers} B.~W.,   {McLaughlin}
  M.~A.,  2011, \mn@doi [\mnras] {10.1111/j.1365-2966.2011.18917.x}, \href
  {http://adsabs.harvard.edu/abs/2011MNRAS.415.3065K} {415, 3065}

\bibitem[\protect\citeauthoryear{{Keane} et~al.,}{{Keane}
  et~al.}{2016}]{2016Natur.530..453K}
{Keane} E.~F.,  et~al., 2016, \mn@doi [\nat] {10.1038/nature17140}, \href
  {http://adsabs.harvard.edu/abs/2016Natur.530..453K} {530, 453}

\bibitem[\protect\citeauthoryear{{Keane} et~al.,}{{Keane}
  et~al.}{2018}]{2018MNRAS.473..116K}
{Keane} E.~F.,  et~al., 2018, \mn@doi [\mnras] {10.1093/mnras/stx2126}, \href
  {http://adsabs.harvard.edu/abs/2018MNRAS.473..116K} {473, 116}

\bibitem[\protect\citeauthoryear{{Kolmogorov}}{{Kolmogorov}}{1933}]{kolmogorov}
{Kolmogorov} A.,  1933, G.\ Ist.\ Ital.\ Attuari.\, 4, 83

\bibitem[\protect\citeauthoryear{{Lawrence}, {Vander Wiel}, {Law}, {Burke
  Spolaor}  \& {Bower}}{{Lawrence} et~al.}{2017}]{2017AJ....154..117L}
{Lawrence} E.,  {Vander Wiel} S.,  {Law} C.,  {Burke Spolaor} S.,   {Bower}
  G.~C.,  2017, \mn@doi [\aj] {10.3847/1538-3881/aa844e}, \href
  {http://adsabs.harvard.edu/abs/2017AJ....154..117L} {154, 117}

\bibitem[\protect\citeauthoryear{{Lorimer}, {Bailes}, {McLaughlin}, {Narkevic}
  \& {Crawford}}{{Lorimer} et~al.}{2007}]{2007Sci...318..777L}
{Lorimer} D.~R.,  {Bailes} M.,  {McLaughlin} M.~A.,  {Narkevic} D.~J.,
  {Crawford} F.,  2007, \mn@doi [Science] {10.1126/science.1147532}, \href
  {http://adsabs.harvard.edu/abs/2007Sci...318..777L} {318, 777}

\bibitem[\protect\citeauthoryear{{Macquart} \& {Ekers}}{{Macquart} \&
  {Ekers}}{2018a}]{2018MNRAS.tmp.1976M}
{Macquart} J.-P.,  {Ekers} R.~D.,  2018a, \mn@doi [\mnras]
  {10.1093/mnras/sty2083}, \href
  {http://adsabs.harvard.edu/abs/2018MNRAS.tmp.1976M} {}

\bibitem[\protect\citeauthoryear{{Macquart} \& {Ekers}}{{Macquart} \&
  {Ekers}}{2018b}]{2018MNRAS.474.1900M}
{Macquart} J.-P.,  {Ekers} R.~D.,  2018b, \mn@doi [\mnras]
  {10.1093/mnras/stx2825}, \href
  {http://adsabs.harvard.edu/abs/2018MNRAS.474.1900M} {474, 1900}

\bibitem[\protect\citeauthoryear{{Macquart} et~al.,}{{Macquart}
  et~al.}{2010}]{2010PASA...27..272M}
{Macquart} J.-P.,  et~al., 2010, \mn@doi [\pasa] {10.1071/AS09082}, \href
  {http://adsabs.harvard.edu/abs/2010PASA...27..272M} {27, 272}

\bibitem[\protect\citeauthoryear{{Macquart}, {Shannon}, {Bannister}, {James},
  {Ekers}  \& {Bunton}}{{Macquart} et~al.}{2018}]{jpnew}
{Macquart} J.-P.,  {Shannon} R.~M.,  {Bannister} K.~W.,  {James} C.~W.,
  {Ekers} R.~D.,   {Bunton} J.~D.,  \noop{3001}Submitted to ApJ Letters, 2018

\bibitem[\protect\citeauthoryear{{Murdoch}, {Crawford}  \& {Jauncey}}{{Murdoch}
  et~al.}{1973}]{1973ApJ...183....1M}
{Murdoch} H.~S.,  {Crawford} D.~F.,   {Jauncey} D.~L.,  1973, \mn@doi [\apj]
  {10.1086/152202}, \href {http://adsabs.harvard.edu/abs/1973ApJ...183....1M}
  {183, 1}

\bibitem[\protect\citeauthoryear{{Oppermann}, {Connor}  \& {Pen}}{{Oppermann}
  et~al.}{2016}]{2016MNRAS.461..984O}
{Oppermann} N.,  {Connor} L.~D.,   {Pen} U.-L.,  2016, \mn@doi [\mnras]
  {10.1093/mnras/stw1401}, \href
  {http://adsabs.harvard.edu/abs/2016MNRAS.461..984O} {461, 984}

\bibitem[\protect\citeauthoryear{{Oslowski} et~al.,}{{Oslowski}
  et~al.}{2018a}]{2018ATel11385....1O}
{Oslowski} S.,  et~al., 2018a, The Astronomer's Telegram, \href
  {http://adsabs.harvard.edu/abs/2018ATel11385....1O} {11385}

\bibitem[\protect\citeauthoryear{{Oslowski} et~al.,}{{Oslowski}
  et~al.}{2018b}]{2018ATel11396....1O}
{Oslowski} S.,  et~al., 2018b, The Astronomer's Telegram, \href
  {http://adsabs.harvard.edu/abs/2018ATel11396....1O} {11396}

\bibitem[\protect\citeauthoryear{{Oslowski} et~al.,}{{Oslowski}
  et~al.}{2018c}]{2018ATel11851....1O}
{Oslowski} S.,  et~al., 2018c, The Astronomer's Telegram, \href
  {http://adsabs.harvard.edu/abs/2018ATel11851....1O} {11851}

\bibitem[\protect\citeauthoryear{{Petroff} et~al.,}{{Petroff}
  et~al.}{2014}]{2014ApJ...789L..26P}
{Petroff} E.,  et~al., 2014, \mn@doi [\apjl] {10.1088/2041-8205/789/2/L26},
  \href {http://adsabs.harvard.edu/abs/2014ApJ...789L..26P} {789, L26}

\bibitem[\protect\citeauthoryear{{Petroff} et~al.,}{{Petroff}
  et~al.}{2015}]{2015MNRAS.447..246P}
{Petroff} E.,  et~al., 2015, \mn@doi [\mnras] {10.1093/mnras/stu2419}, \href
  {http://adsabs.harvard.edu/abs/2015MNRAS.447..246P} {447, 246}

\bibitem[\protect\citeauthoryear{{Petroff} et~al.,}{{Petroff}
  et~al.}{2016}]{2016PASA...33...45P}
{Petroff} E.,  et~al., 2016, \mn@doi [\pasa] {10.1017/pasa.2016.35}, \href
  {http://adsabs.harvard.edu/abs/2016PASA...33...45P} {33, e045}

\bibitem[\protect\citeauthoryear{{Petroff} et~al.,}{{Petroff}
  et~al.}{2017}]{2017MNRAS.469.4465P}
{Petroff} E.,  et~al., 2017, \mn@doi [\mnras] {10.1093/mnras/stx1098}, \href
  {http://adsabs.harvard.edu/abs/2017MNRAS.469.4465P} {469, 4465}

\bibitem[\protect\citeauthoryear{{Price} et~al.,}{{Price}
  et~al.}{2018}]{2018ATel11376....1P}
{Price} D.~C.,  et~al., 2018, The Astronomer's Telegram, \href
  {http://adsabs.harvard.edu/abs/2018ATel11376....1P} {11376}

\bibitem[\protect\citeauthoryear{{Ravi} et~al.,}{{Ravi}
  et~al.}{2016}]{2016Sci...354.1249R}
{Ravi} V.,  et~al., 2016, \mn@doi [Science] {10.1126/science.aaf6807}, \href
  {http://adsabs.harvard.edu/abs/2016Sci...354.1249R} {354, 1249}

\bibitem[\protect\citeauthoryear{{Schmidt}}{{Schmidt}}{1968}]{1968ApJ...151..393S}
{Schmidt} M.,  1968, \mn@doi [\apj] {10.1086/149446}, \href
  {http://adsabs.harvard.edu/abs/1968ApJ...151..393S} {151, 393}

\bibitem[\protect\citeauthoryear{{Shannon} et~al.,}{{Shannon}
  et~al.}{2017}]{2017ATel11046....1S}
{Shannon} R.~M.,  et~al., 2017, The Astronomer's Telegram, \href
  {http://adsabs.harvard.edu/abs/2017ATel11046....1S} {11046}

\bibitem[\protect\citeauthoryear{{Shannon} et~al.,}{{Shannon}
  et~al.}{2018}]{craft_nature}
{Shannon} R.~M.,  et~al., 2018, \mn@doi [Nature] {10.1038/s41586-018-0588-y}

\bibitem[\protect\citeauthoryear{{Smirnov}}{{Smirnov}}{1948}]{smirnov}
{Smirnov} N.,  1948, \mn@doi [Annals of Mathematical Statistics]
  {doi:10.1214/aoms/1177730256}, 19, 279

\bibitem[\protect\citeauthoryear{{Spitler} et~al.,}{{Spitler}
  et~al.}{2016}]{2016Natur.531..202S}
{Spitler} L.~G.,  et~al., 2016, \mn@doi [\nat] {10.1038/nature17168}, \href
  {http://adsabs.harvard.edu/abs/2016Natur.531..202S} {531, 202}

\bibitem[\protect\citeauthoryear{{Tendulkar} et~al.,}{{Tendulkar}
  et~al.}{2017}]{2017ApJ...834L...7T}
{Tendulkar} S.~P.,  et~al., 2017, \mn@doi [\apjl] {10.3847/2041-8213/834/2/L7},
  \href {http://adsabs.harvard.edu/abs/2017ApJ...834L...7T} {834, L7}

\bibitem[\protect\citeauthoryear{{The CHIME/FRB Collaboration} et~al.,}{{The
  CHIME/FRB Collaboration} et~al.}{2018}]{2018ApJ...863...48T}
{The CHIME/FRB Collaboration} et~al., 2018, \mn@doi [\apj]
  {10.3847/1538-4357/aad188}, \href
  {http://adsabs.harvard.edu/abs/2018ApJ...863...48T} {863, 48}

\bibitem[\protect\citeauthoryear{{Thornton} et~al.,}{{Thornton}
  et~al.}{2013}]{2013Sci...341...53T}
{Thornton} D.,  et~al., 2013, \mn@doi [Science] {10.1126/science.1236789},
  \href {http://adsabs.harvard.edu/abs/2013Sci...341...53T} {341, 53}

\bibitem[\protect\citeauthoryear{{Vedantham}, {Ravi}, {Hallinan}  \&
  {Shannon}}{{Vedantham} et~al.}{2016}]{2016ApJ...830...75V}
{Vedantham} H.~K.,  {Ravi} V.,  {Hallinan} G.,   {Shannon} R.~M.,  2016,
  \mn@doi [\apj] {10.3847/0004-637X/830/2/75}, \href
  {http://adsabs.harvard.edu/abs/2016ApJ...830...75V} {830, 75}

\bibitem[\protect\citeauthoryear{{Wall}}{{Wall}}{1996}]{1996IAUS..175..547W}
{Wall} J.~V.,  1996, in {Ekers} R.~D.,  {Fanti} C.,   {Padrielli} L.,  eds,
  IAU Symposium Vol. 175, Extragalactic Radio Sources. p.~547

\bibitem[\protect\citeauthoryear{{Wall}, {Pearson}  \& {Longair}}{{Wall}
  et~al.}{1980}]{1980MNRAS.193..683W}
{Wall} J.~V.,  {Pearson} T.~J.,   {Longair} M.~S.,  1980, \mn@doi [\mnras]
  {10.1093/mnras/193.3.683}, \href
  {http://adsabs.harvard.edu/abs/1980MNRAS.193..683W} {193, 683}

\bibitem[\protect\citeauthoryear{Wilks}{Wilks}{1938}]{wilks1938}
Wilks S.~S.,  1938, \mn@doi [Ann. Math. Statist.] {10.1214/aoms/1177732360}, 9,
  60

\makeatother
\end{thebibliography}

\appendix

\section{Explicit beam-dependence of telescope sensitivity}
\label{sec:extended}

The abstract definition of the parameter set $\bmath{\theta}$ in section~\ref{sec:derivation} makes it difficult to relate to specific experimental examples. Below is explicitly outlined its application in the context of a ``completeness limit'' defined by observable fluence (i.e.\ reduced by the telescope beam), as discussed in section~\ref{sec:completeness}, and used by \citet{2018MNRAS.474.1900M} and \citet{2018MNRAS.475.1427B}.

Let the beamshape of a telescope be described by the direction-dependent relative sensitivity $B(\Omega)$, such that $B(\Omega) \le 1$, with (in general) equality at beam centre. Ignoring frequency dependencies (which could be treated within the set $\bmath{\theta}$), an FRB with fluence $F$ would be registered with modified fluence $F^{\prime}$:
\begin{eqnarray}
F^{\prime}(B) & = & B F \label{eq:fprime}.
\end{eqnarray}
The rate of FRBs, and hence the source term $k$ of equation~(\ref{eq:Kk}), will not depend on $B$, and FRBs will arrive uniformly in solid-angle $\Omega$. Furthermore, assume that the threshold $F_{\rm th}(\bmath{\theta})$ depends on $B$ only via equation~(\ref{eq:fprime}). It is useful to explicitly remove $B$ from the parameter set $\bmath{\theta}$ by defining:
\begin{eqnarray}
\bmath{\theta} & = & \left\{ \bmath{\phi},B \right\},
\end{eqnarray}
i.e.\ $\phi$ is the set of all confounding parameters except $B$. Hence, we can define:
\begin{eqnarray}
F_{\rm th}(\bmath{\theta}) & = & \frac{F_{\rm th}(\bmath{\phi})}{B},
\end{eqnarray}
and, since the event rate is not dependent on the beam, it can be written:
\begin{eqnarray}
R(\bmath{\theta}) & = & R(\bmath{\phi}) \nonumber \\
& = & k(\bmath{\phi}) \left(\frac{F}{F_0}\right)^{\alpha}.
\end{eqnarray}
The total integral $C$ over all parameter space (c.f.\ equation~(\ref{eq:cnorm})) can be calculated as:
\begin{eqnarray}
C = \int  d\bmath{\phi} \int d \Omega \int_{F_{\rm th}(\bmath{\theta})}^{F_m(\bmath{\theta})} k(\bmath{\phi}) \frac{-\alpha}{F_0} \left(\frac{F}{F_0}\right)^{\alpha-1} dF. \label{eq:bnorm}
\end{eqnarray}
This can also be written as an integral over beam-factor $B$, by defining the solid angle viewed at each value of $B$, $\Omega(B)$ (see \citet{ME}). Hence, $C$ becomes:
\begin{eqnarray}
C & = & \int k(\bmath{\phi}) d\bmath{\phi} \int  \Omega(B) dB \int_{F_{\rm th}(\bmath{\phi})/B}^{F_m(\bmath{\phi})/B}  \nonumber \\
&& \frac{-\alpha}{F_0} \left( \frac{F}{F_0} \right)^{\alpha-1} dF \label{eq:bint1}
\end{eqnarray}
As with equation~(\ref{eq:cnorm}), the integration over $F$ in equation~(\ref{eq:bint1}) can be performed explicitly:
\begin{eqnarray}
C & = & \int k(\bmath{\phi}) d\bmath{\phi} \int  \Omega(B) dB \nonumber \\
&& \left[ \left(\frac{F_{\rm th}(\bmath{\phi})}{B F_0}\right)^{\alpha} - \left(\frac{F_{m}(\bmath{\phi})}{B F_0}\right)^{\alpha}  \right]. \label{eq:bnorm2}
\end{eqnarray}
Changing the definitions of relative minimum and maximum fluences $s$ and $b$ in equation~(\ref{eq:relative}) to be relative to observed values:
\begin{eqnarray}
s & = & B \frac{F}{F_{\rm th}(\bmath{\phi})} \nonumber \\
b & = & B \frac{F}{F_{\rm max}(\bmath{\phi})},
\end{eqnarray}
and inserting these into equation~(\ref{eq:bnorm2}) with $ds = B F_{\rm th}^{-1} dF$, produces:
\begin{eqnarray}
C & = & \int k(\bmath{\phi}) d\bmath{\phi} \int  \Omega(B) dB \cdot \nonumber \\
&& \int_{s=1}^{b}\frac{-\alpha}{F_0} \left( \frac{s F_{\rm th}}{B F_0} \right)^{\alpha-1} \frac{F_{\rm th}}{B} ds. \label{eq:bnorm3}
\end{eqnarray}
What we wish to calculate is the normalised probability $P(s)$, since neither $B$ nor $\bmath{\phi}$ are observable. This is defined as:
\begin{eqnarray}
P(s) & = & \int d\phi \int dB P(s,B,\phi),
\end{eqnarray}
where the joint probability $P(s,B,\phi)$ is simply the normalised negative of the integrand in equation~(\ref{eq:bnorm3}):
\begin{eqnarray}
P(s,B,\phi) & = & \frac{-1}{C} k(\bmath{\phi}) \Omega(B) \alpha s^{\alpha-1} \left( \frac{F_{\rm th}}{B F_0} \right)^{\alpha}.
\end{eqnarray}
Observe that the integrals over $\bmath{\phi}$ and $B$ can be separated out:
\begin{eqnarray}
P(s) & = &\frac{-1}{C} \alpha s^{\alpha-1} F_0^{-\alpha} \cdot \nonumber \\
&&\left[ \int k(\bmath{\phi}) F_{\rm th}^{\alpha} d\phi \right] \left[ \int \Omega(B) B^{-\alpha} dB \right]. \label{eq:ps2}
\end{eqnarray}
Returning to the constant $C$, performing the integration over $s$ in equation~(\ref{eq:bnorm3}) produces:
\begin{eqnarray}
C = \int k(\bmath{\phi}) d\bmath{\phi} \int  \Omega(B) dB \left( \frac{F_{\rm th}}{B F_0} \right)^{\alpha} \left[ 1 - b^{\alpha} \right].
\end{eqnarray}
Noting that $F_{\rm th}$ is a function of $\bmath{\phi}$, but not $B$, and again assuming that $b$ is independent of both parameters, $C$ can be written:
\begin{eqnarray}
C = \left[ 1-b^{\alpha} \right] F_0^{-\alpha} \int  \Omega(B) B^{-\alpha} dB  \int k(\bmath{\phi}) F_{\rm th}^{\alpha}(\bmath{\phi}) d\bmath{\phi}. \label{eq:bnorm4}
\end{eqnarray}
The integrals over $B$ and $\phi$ in equation~(\ref{eq:bnorm4}) for $C$ are identical to, and cancel with, the same integrals in equation~(\ref{eq:ps2}). Removing also the common power of $F_0^{-\alpha}$ produces:
\begin{eqnarray}
P(s) & = & -\alpha \frac{s^{\alpha-1}}{1-b^{\alpha}}
\end{eqnarray}
This is identical to equation~(\ref{eq:ps}), with $s$ and $b$ defined in terms of observable (i.e.\ beam-affected) quantities. All results thus follow, i.e.\ the results of \citet{1970ApJ...162..405C} outlined in section~\ref{sec:theory} hold.

We also note that applying an artificial cut (or completeness limit) $F_c>F_{\rm th}$ in the integrals over $d\bmath{\phi}$, and defining $s$ and $b$ relative to $F_c$ rather than $F_{\rm th}$, produces an identical result. Therefore, the application of \citet{1970ApJ...162..405C} by \citet{2018MNRAS.474.1900M} and \citet{2018MNRAS.475.1427B} to beam-affected FRB fluences above a completeness limit is valid.

% Don't change these lines
\bsp	% typesetting comment
\label{lastpage}
\end{document}